\documentclass[aps,onecolumn,pre]{revtex4}[12]
\usepackage{latexsym, verbatim}
\usepackage{amsmath}
\usepackage{mathrsfs}
\usepackage{amsthm}
\usepackage{amssymb}
\usepackage{dsfont}
\usepackage{fancybox}
\usepackage{bm}
\usepackage{fancyhdr}
\usepackage{subfigure}
\usepackage{graphicx}
\usepackage{color}

\newcommand{\cP}{\ensuremath{\mathcal{P}}}
\newcommand{\cT}{\ensuremath{\mathcal{T}}}

\begin{document}


\title{\bf Eigenstates and instabilities of chains with embedded defects}

\author{J. D'Ambroise}
\affiliation{Department of Mathematics,
Bard College,
Annandale-on-Hudson, NY 12504, USA}

\author{P.G.~Kevrekidis}
\affiliation{Department of Mathematics and Statistics, University of Massachusetts,
Amherst, Massachusetts 01003-4515, USA}

\author{S. Lepri}
\affiliation{CNR-Consiglio Nazionale delle Ricerche, Istituto dei Sistemi
Complessi, via Madonna del piano 10, I-50019 Sesto Fiorentino, Italy}

\begin{abstract}
We consider the eigenvalue problem for one-dimensional
linear Schr{\"o}dinger lattices (tight-binding) with an
embedded few-sites linear or nonlinear,
Hamiltonian or non-conservative defect (an oligomer). Such a problem arises
when considering scattering states in the presence of (generally
complex) impurities as well as in the stability analysis
of nonlinear waves.
We describe a general approach based on a matching of
solutions of the linear portions of the lattice  at the location of the
oligomer defect. As specific examples we discuss both linear and
nonlinear, Hamiltonian and $\cP \cT$-symmetric dimers and trimers.
In the linear case, this approach provides us a handle for
semi-analytically computing the spectrum [this amounts to the
solution of a polynomial equation]. In the nonlinear case, it enables
the computation of the linearization spectrum around the stationary
solutions. The calculations showcase the oscillatory instabilities
that strongly nonlinear
states typically manifest.
\end{abstract}

\maketitle

\begin{quotation}
{\bf
We consider the time evolution of a quantum mechanical wave function as governed by the Schr\"odinger equation.  The wave function is distributed spatially on a discrete one-dimensional lattice, i.e. a chain of nodes indexed by integers, so that the spatial derivatives are replaced by differences.  The potential function is nonzero only at a few center sites on the lattice, representing either physical impurities or other obstacles such as an external field
or a nonlinear material.  Since the general solution of the zero potential
problem is well-known, we begin by constructing it
on the outer (left and right) portions of the lattice.  Working our way
toward the impurity sites using the restraints of the discrete Schr\"odinger equation, we
find that appropriately defined portions of the outer solution must satisfy a polynomial equation.  Our method is also applied to the (again) discrete but
(now) nonlinear Schr\"odinger equation.  Here known stationary solutions are
acted upon by a time-dependent perturbation, and we find that appropriately defined portions of the perturbation must satisfy polynomial equations.  The main point is to show that the polynomial conditions we derive accurately determine the dynamical stability of the solutions.  The success of our method in tracking
the associated linear and nonlinear spectra is presented throughout the
linear and nonlinear cases.  In order to demonstrate the generality of our approach we show examples using both real valued Hamiltonians and purely complex parity-time symmetric potentials.
}
\end{quotation}

\section{Introduction}

Linear lattices with embedded impurities
have been considered to model many different physical systems.
A familiar example is the case of
mass defects in an otherwise pure harmonic crystal.
In the context of the tight-binding description of
electron transport, the nonlinear terms describe the strong
interaction with local vibrations at the impurity site ~\cite{tsir1}.
Models of the same type have been used to describe tunneling
through a magnetic impurity connected to two perfect leads in the
presence of a magnetic field~\cite{moli}.
Another wide domain of application is the one of nonlinear optics.
The type of system thereby considered is an array of coupled
waveguides consisting of (at least) two types of materials
one linear and one of Kerr type, see e.g. ~\cite{efrem}.
Several problems have been addressed, ranging from the existence
of localized solutions residing on the nonlinear portion (nonlinear
impurity modes \cite{kovalev}) to the scattering and transmission of
plane waves through it \cite{Hennig99}.
{ A model which is often used  to
describe such a wide palette of problems is the
discrete nonlinear Schr\"odinger equation. There, the
discretization of the continuum Schr\"odinger operator
is emulated by a predominant nearest-neighbor interaction.
As is known,
the seminal works of Davydov \cite{davydov} and Holstein~\cite{holstein}
were an early stimulus yielding a considerable body of
work on discrete solitary waves (by pioneers
such as Bishop, Campbell, Scott, and others); see e.g. the more
recent account of \cite{Scott2003}.
Genuine nonlinear features, like the self-trapping transitions, are already
present in few-degrees-of-freedom system, like for instance
the dimer \cite{kenkre}. Early on in this literature, problems with
one or multiple defect nodes embedded in linear chains were considered,
see e.g. the work of~\cite{molina1,molina2} and references therein. }

An important question concerns the spectral characteristics of the problem
in the linear regime and the dynamical stability of the solutions in the
nonlinear regime (i.e., when the embedded impurities are
genuinely nonlinear). To the best of our knowledge no systematic study of
such an important issue been presented in the literature.
In Ref.~\cite{malaz} the continuum case of Nonlinear Schr\"odinger equation with
a localized ($\delta$-function) nonlinearity was considered.
For discrete lattices, {similar techniques have been used to match an ansatz on the linear part of the lattice with the relevant Schr\"odinger equation condition on the nonlinear site(s) \cite{BMalDing, BMalBrazhnyi, BMalBrazhnyiSSM}.  In \cite{BMalDing, SolPinHS} ``hot spot" sites carrying linear gain and cubic nonlinearity on an infinite lattice were investigated within a lossy
bulk medium.  A}n analysis of a related problem, the nonlinear Fano effect, has been reported in \cite{Miroshnichenko2009} (see also references therein).
Another related reference is \cite{Tietsche2008} where
a bifurcation analysis of an open chain with nonlinearity and
disorder is performed.

The aim of the present paper is to formulate (in general) and to
solve (in some specific cases) the spectral problem for a discrete Schr\"odinger equation
with $N{\geq 2}$ embedded defects (an $N$-site oligomer \cite{pgk}). Equations of this type
emerge in, at least, two interesting cases. The first
is the one of linear chains with localized \textit{complex} on-site potentials.
The second one arises when considering
the stability problem of complex (i.e. current carrying)
scattering solutions of the discrete nonlinear Schr\"odinger equation \cite{Hennig99}.
Such solutions can be computed exactly by the transfer matrix method
\cite{Delyon86,Wan90,Li96,Hennig99} and linearization around
them yields linear equations of the same type.

Our motivation to approach this issue is twofold. The first is to study
the scattering problem by open systems possessing a
$\cP \cT$-symmetry~\cite{R1,R2,R21}. This theme is rapidly
evolving into a major area of research partly due to its
providing an intriguing alternative set of non-Hermitian
Hamiltonians with potentially real eigenvalues and partly due to
its experimental realizations in the field of nonlinear
optics, both in the case of dimers~\cite{dnc1,dnc2}
and even in that of whole lattices~\cite{dnc_nat}.
The second motivation stems from a recent study
of non-reciprocal wave propagation through non-mirror symmetric
nonlinear lattices \cite{lepri}. These are a minimal model for
a ``wave diode'', namely for devices capable to selectively rectify
wave energy through a nonlinear medium of propagation. Relevant
applications arise for instance for phonon scattering at a nonlinear interface layer between two
very dissimilar crystals \cite{kosevich}, acoustic waves in sonic materials
\cite{chiara1} or in the so-called all-optical diode for photonic
applications~\cite{l5}.

The solution of infinite-dimensional problems like the above
is technically more
involved than a straightforward matrix diagonalization. Indeed,
to solve the problem exactly one should impose
the solution a definite plane wave form (with generally complex wave numbers) in
the two semi-infinite linear parts of the chain. The matching of such waves
through the oligomer portion reduces the infinite-dimensional problem
to an homogeneous linear system of $2N$ equations, whose solvability
condition, along with the dispersion relations, yields a set of nonlinear
equations for the unknowns. { We characterize our method as
``semi-analytic", namely it is exact yet it relies on numerical solutions of a
high degree polynomial equation. As a relevant disclaimer here, we should
point out that our method is not proposed as an advantageous computational
alternative to standard full eigenvalue solvers.
Instead, our results are intended to
give mathematical form and physical understanding,
as close to analytic as is possible, to the output of an
otherwise purely numerical computation. For instance, the process 
provides insight on the functional form of the corresponding eigenvectors.}

The details of this method  will be
presented for a linear $\cP \cT$-symmetric case~\cite{R1,R2,R21}. In the
latter, the imposition of the parity ($\cP$, associated with spatial
reflection)
and time-reversal ($\cT$, associated with temporal reflection
and also $i \rightarrow -i$) symmetries
leads to an imaginary (growth/decay) part of the Schr{\"o}dinger
potential which is spatially odd. The relevant more general
constraint, in fact, reads $V(x)=V^{\star}(-x)$ (where $\star$ stands
for complex conjugation); i.e., the real part of the relevant
potential needs to be even. This setting
will be compared/contrasted to the Hamiltonian case, where the
potential is real.

The paper is organized as follows. In section \ref{sec: L}, we will
discuss the simplest case namely that of linear oligomers.
Although simple, this  case is useful to set the stage of
the general methodology
and to illustrate some basic features that will be useful throughout
the rest of the paper.
In section \ref{sec: N}, we turn to the nonlinear case. In each of these,
we first present the relevant analysis and subsequently we corroborate
it on the basis of numerical computations. Finally, in section~\ref{conc},
we present a summary of our findings and some potential directions for future
study.

\section{Linear Case}
\label{sec: L}

In this section we present the strategy to compute the spectrum and eigenstates
of a general linear problem where $N$-site complex defects are embedded in
an otherwise
homogeneous lattice. More precisely, we consider the dynamic evolution
of a one-dimensional chain governed by the linear discrete
Schr\"odinger equation
\begin{equation}
\label{eq: TDDLS}
i\dot\phi_n-V_n\phi_n+\phi_{n+1}+\phi_{n-1}=0
\end{equation}
where $\phi_n(t) \in\mathds{C}$.
The complex potential $V_n\in\mathds{C}$ is zero everywhere except
for $1\leq n\leq N$ for some integer $N\geq 2$.

We look for solutions of the form $\phi_n(t)=a_n e^{i \omega t}$
so that (\ref{eq: TDDLS}) becomes the time-independent condition
\begin{equation}
\label{eq: DLS}
 a_{n+1}+a_{n-1}-(V_n+\omega) a_n=0.
\end{equation}
Note that $a_n$ is complex as well.

\subsection{Theoretical Analysis}
\label{sec: Lthy}

Our goal is to present a strategy for the analytic computation of
the eigenstates $a_n$ and eigenvalues $\omega$, assuming that $V_n$ has been prescribed.
First we begin with an observation which applies to the portions of the lattice on the
left and right of the $N$ central sites.
The zero potential ($V_n=0$) version of
equation (\ref{eq: DLS}) has a general solution of the form
$a_n=Ae^{i \kappa n}+Be^{-i\kappa n}$ for $A, B \in\mathds{C}$ and $\kappa\in\mathds{C}$
satisfying the dispersion relation $2\cos(\kappa)=\omega$.  Thus we begin with the ansatz
\begin{eqnarray}
a_n&=&
\left\{
\begin{array}{cc}
 Ae^{i\kappa n}+Be^{-i\kappa n} &n< 1\\
Ce^{i\kappa n}+De^{-i\kappa n} & n> N
\end{array}
\right.\label{eq: Lansatz}
\end{eqnarray}
for $A, C, B, D\in\mathds{C}$ and $\kappa$ satisfying
 \begin{equation}
 \omega=z+\frac{1}{z}
 \label{eq: omegaz}
 \end{equation}
  where we use the shorthand notation $z=e^{i\kappa}\neq 0$.  Imposing (\ref{eq: Lansatz}) and (\ref{eq: omegaz}) is enough to satisfy (\ref{eq: DLS}) for all $n$ except $n=0, 1, \dots, N, N+1$.  Applying (\ref{eq: Lansatz}) to (\ref{eq: DLS}) with $n=0, N+1$ shows that the
formula (\ref{eq:  Lansatz}) is applicable at $n=1, N$. Since in practice we deal with a finite length lattice we introduce indexing $n_0\leq n \leq m_0$ for large magnitude integers $n_0<<1$ and $m_0>>N$.  Imposing~\cite{footnote1}
homogeneous Dirichlet boundary conditions $a_{n_0-1}=a_{m_0+1}=0$ shows that  $\omega = a_{n_0+1}/a_{n_0} = a_{m_0-1}/a_{m_0}$.  Combining this with (\ref{eq: Lansatz})  gives  $B =  - A  e^{2i\kappa (n_0-1)}$ and $D = - C  e^{2i\kappa (m_0+1)}$.    Thus we have
\begin{eqnarray}
a_n&=&
\left\{\begin{array}{cc}
 A c_n &n \leq 1\\
C  d_n& n \geq N
\end{array}
\right.   \label{eq: Lextansatz}
\end{eqnarray}
for $ c_n =   z^n - z^{ 2n_0-2-n}$ and $d_n=z^{n}  -  z^{2m_0+2-n}$.

At this point the imposition of (\ref{eq: Lextansatz}) and (\ref{eq: omegaz}) is enough to satisfy all but the $N$  equations in (\ref{eq: DLS}) with $n=1, \dots, N$.  In the following sections we outline strategies for completing the computation of $a_n$ and $\omega$ in the cases of embedded few-site defects
and apply the method to our $\cP \cT$-symmetric
or Hamiltonian oligomer.

\subsubsection{\bf \emph{Oligomer of length two}}
\label{sec: Lthy2}

For $N=2$ we examine the two equations (\ref{eq: DLS}) for $n=1,2$ which by (\ref{eq: Lextansatz})  can be written as
\begin{equation}
P
\left[
\begin{array}{c}
a_0\\
a_1\\
a_2\\
a_{3}
\end{array}
\right]
=
P
Q
\left[
\begin{array}{c}
A\\
C
\end{array}
\right]
=0
\mbox{\quad for \quad}
P=
\left[
\begin{array}{cccc}
1 & -(V_1+\omega) & 1 & 0\\
0 & 1  & -(V_2+\omega) & 1
\end{array}
\right]
\mbox{ \quad and \quad }
Q=
\left[
\begin{array}{cc}
c_0& 0\\
c_1 & 0\\
0 & d_2 \\
0 & d_3
\end{array}
\right].
\label{eq: P2Q2}
\end{equation}
 In other words, we now have a $2\times 2$ system with unknowns $A, C$ and coefficient matrix $PQ$.  Imposing zero determinant $det(PQ)=0$ gives a solvability condition, which by (\ref{eq: omegaz}) amounts to requiring that allowable $z$ are roots of a polynomial with coefficients in terms of $V_n$.    For each root $z$ one computes the corresponding $A, C$ by the equation in (\ref{eq: P2Q2}) so that finally an eigenvalue $\omega$ and eigenvector $a_n$ are known by (\ref{eq: omegaz}) and (\ref{eq: Lextansatz}).  Note that roots $z=\pm 1$ correspond to $c_n=d_n=a_n=0$ and should be disregarded so as to obtain non-trivial eigenvectors.

As an example, consider the special case $V_1+V_2=0$.  One can compute by hand that the determinant condition, when expressed in terms of a positive power polynomial,  is
\begin{eqnarray}
0&=& z^{2(m_0-n_0)+6} + (V_1V_2-1)z^{ 2(m_0-n_0)+4}
  - V_1 z^{2m_0+1}
  - V_1V_2 z^{ 2m_0}
 \nonumber\\
&&
   -   V_2 z^{ 2m_0 -1}
 - V_2 z^{-2n_0+7}
 - V_1V_2 z^{ -2n_0+6}
 - V_1  z^{-2n_0+5}
 +(V_1V_2-1) z^{ 2}  +1  \label{eq: lin2V1=-V2polyn}
\end{eqnarray}
for which $z=\pm 1$ are each double roots.  This reduces the number of relevant roots of (\ref{eq: lin2V1=-V2polyn}) to $2(m_0-n_0+1)\equiv 2L$, twice the length of the lattice.  One can easily show, using the symmetry of the coefficients in (\ref{eq: lin2V1=-V2polyn}), that solutions of (\ref{eq: lin2V1=-V2polyn}) occur in reciprocal pairs $z, \frac{1}{z}$.  Combining this with (\ref{eq: omegaz}) shows that our analytic computation indeed yields exactly $L$ eigenvalues $\omega$, counted with multiplicity.

\begin{figure}[tbp]
\begin{center}
\includegraphics[width=8cm,angle=0,clip]{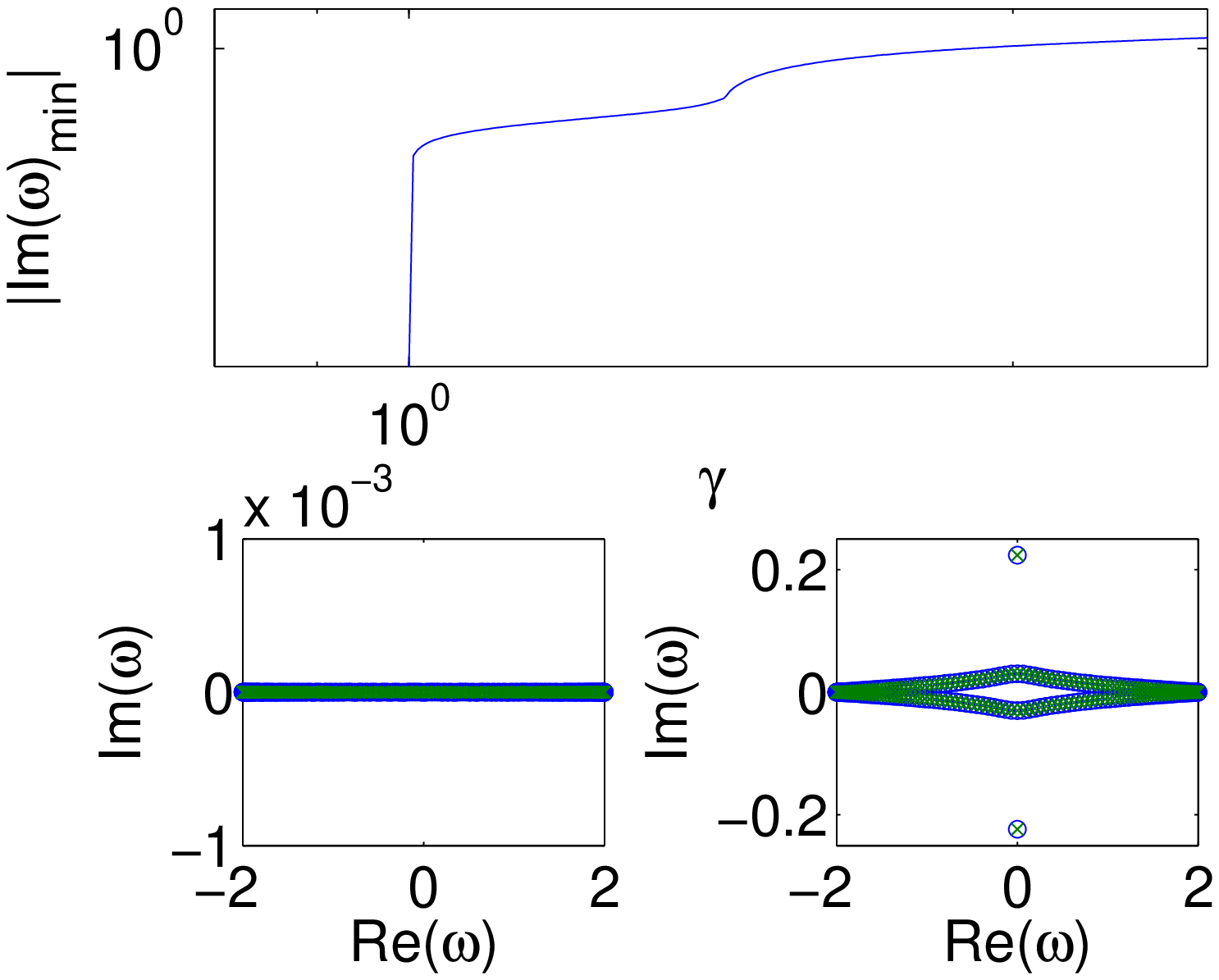}
\includegraphics[width=8cm,angle=0,clip]{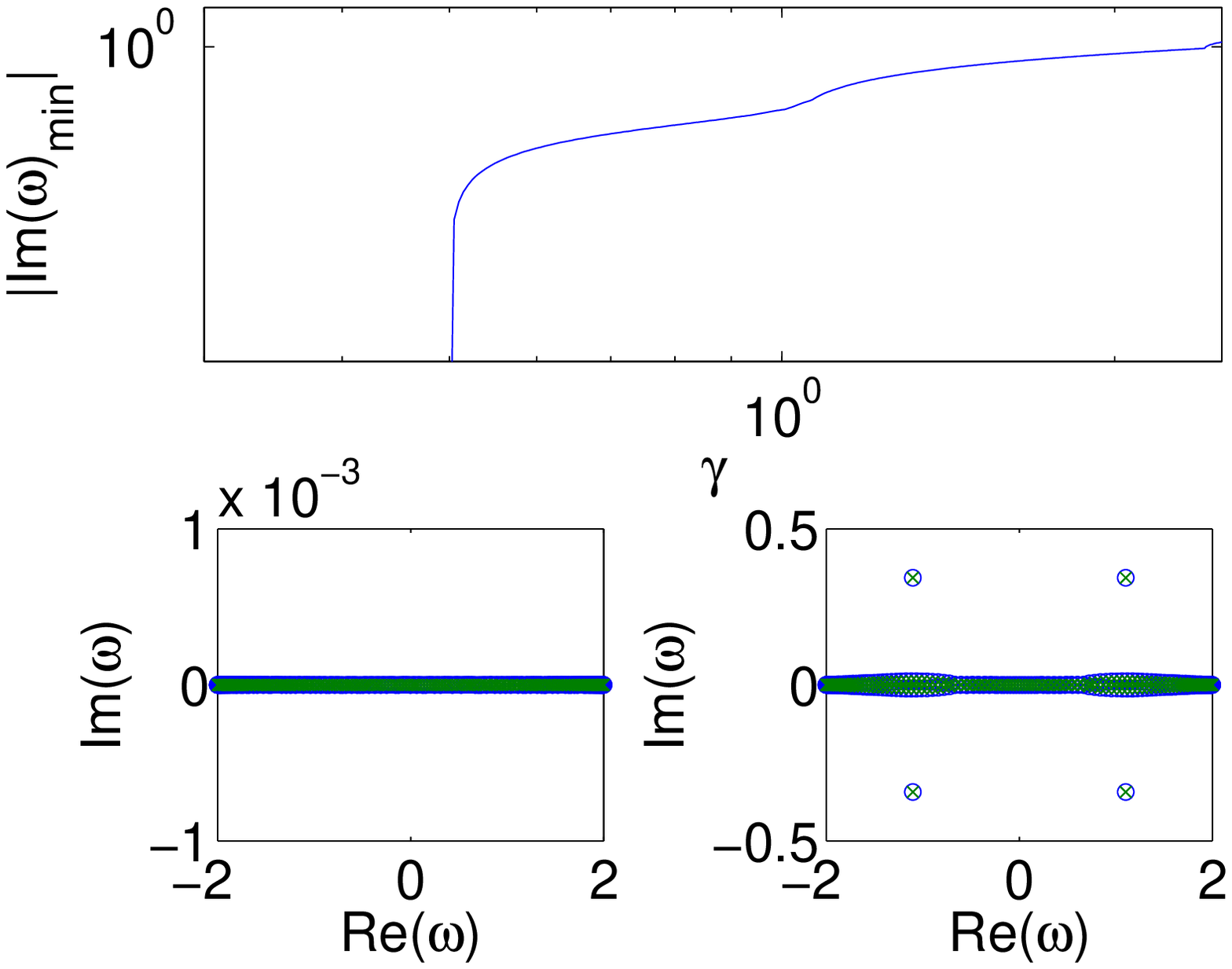}
\end{center}
\caption{
The left set of panels corresponds to the linear $\cP \cT$-symmetric dimer with $N=2$ and $V_1=i\gamma=-V_2$ while the right set of panels to the case of the trimer with $N=3$ and $V_1=i\gamma=-V_3, V_2=0$.  Each set contains a (top) plot of linear stability eigenfrequencies as a function of increasing $\gamma$ as computed an a lattice of length $200$.  The eigenvalue agreement between numeric (circles) and exact (x's) computations  is shown for $\gamma=.5$ (bottom left) and $\gamma=1.5$ (bottom right).
}
\label{lineigvals}
\end{figure}

\begin{figure}[tbp]
\begin{center}
\includegraphics[width=8cm,angle=0,clip]{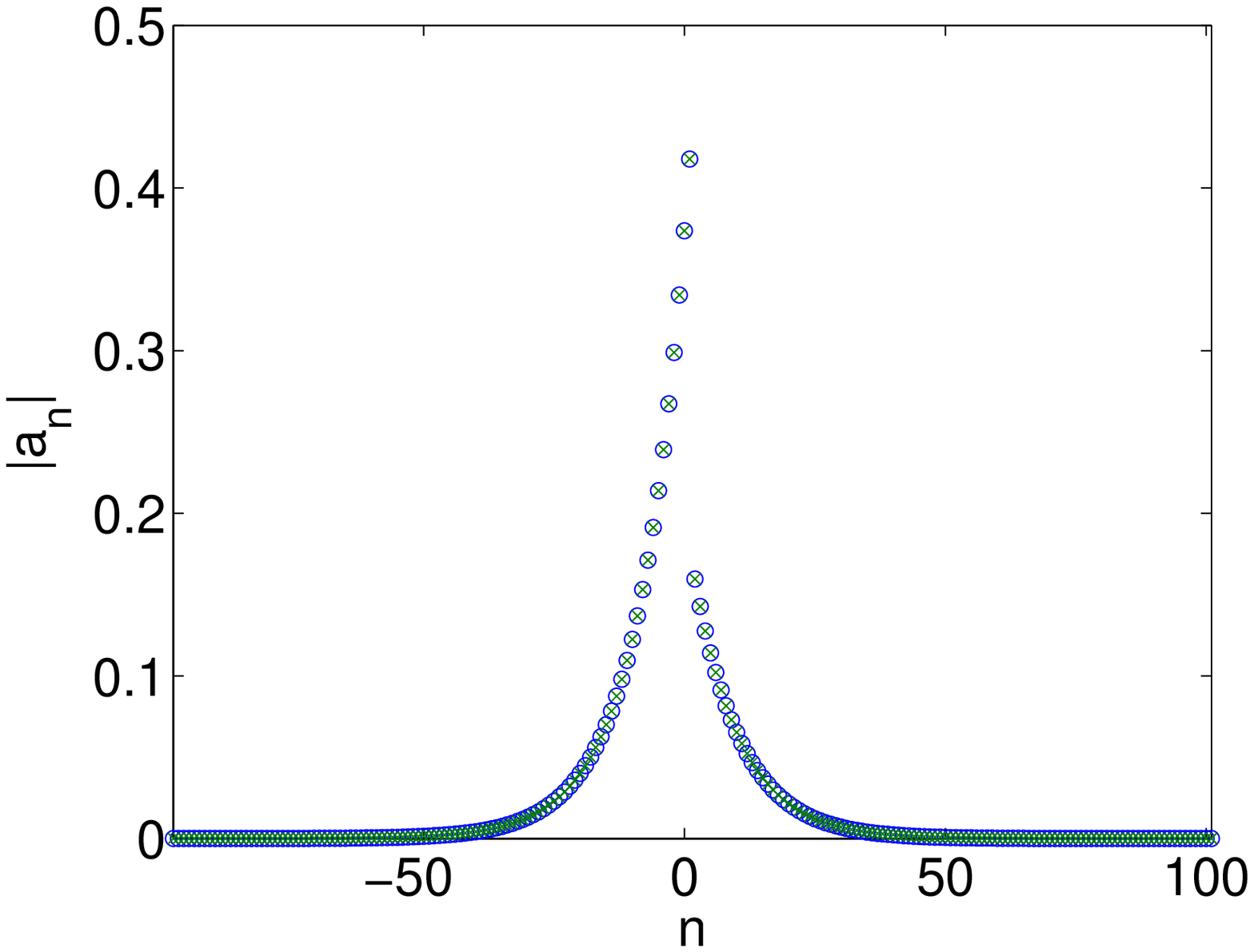}
\includegraphics[width=8cm,angle=0,clip]{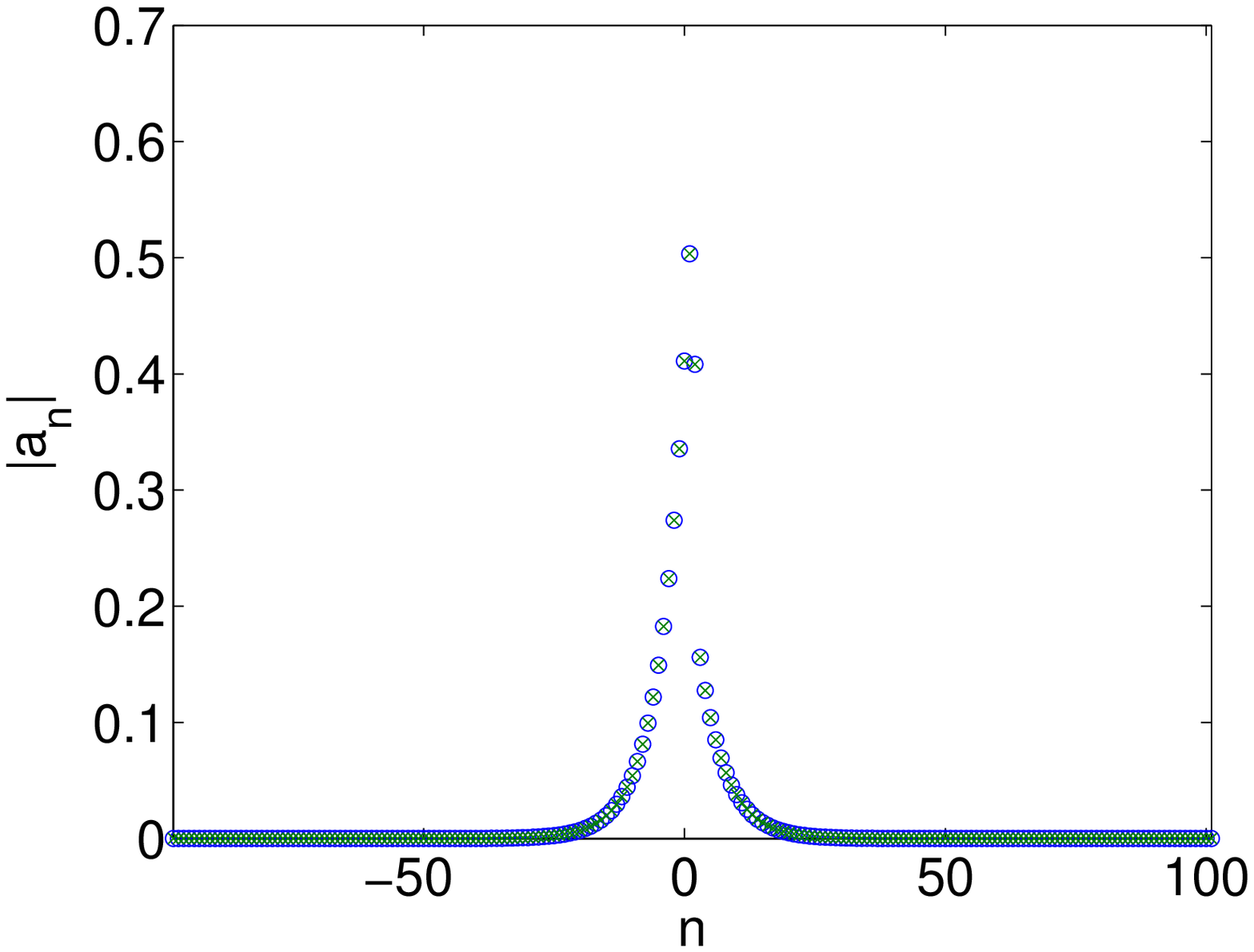}
\end{center}
\caption{
Profiles of modulus of unstable eigenvectors $|a_n|$ are shown for the linear dimer (left) and the linear trimer (right).  These eigenvectors correspond to the eigenvalues with negative imaginary part for $\gamma=1.5$ which are seen in the bottom right panels of Fig. \ref{lineigvals}.  Again, numerical
computations (circles) are shown to agree with the exact computation (x's) and the lattice length is $200$.
}
\label{lineigvects}
\end{figure}

\begin{figure}[tbp]
\begin{center}
\includegraphics[width=8cm,angle=0,clip]{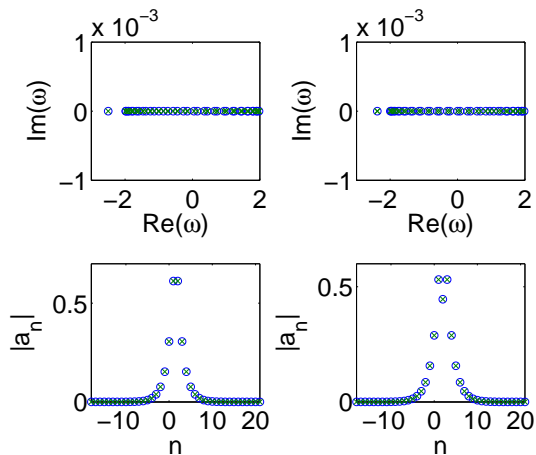}
\end{center}
\caption{
The left set of plots corresponds to the linear Hamiltonian dimer with $N=2$ and $V_1=1=V_2$ while the right set of plots to the case of the linear trimer with $N=3$ and $V_1=1=V_3, V_2=0$.  Each set contains a (top) plot of linear stability eigenvalues and a (bottom) plot of the eigenvector for the defect (point spectrum) mode.  All plots show agreement between the numerical (circles) and semi-analytical (x's) results.  Here the lattice length is $40$.
}
\label{linHameigvalsvects}
\end{figure}

\subsubsection{\bf \emph{Oligomer of length three or greater}}
\label{sec: Lthy>2}

For $N>2$ the strategy is similar to that of the last section but involves an additional step.  First focus on the $N-2$ equations (\ref{eq: DLS}) for $n=2, \dots, N-1$ which we write as
\begin{equation}
\left[
\begin{array}{ccccc}
1 & -(V_2+\omega) &             1                  & & \\
                                 & \ddots & \ddots & \ddots &  \\
  &  &              1             & -(V_{N-1}+\omega)   & 1
\end{array}
\right]
\left[
\begin{array}{c}
a_1\\
\vdots\\
a_N
\end{array}
\right]
=0. \label{eq: LN-2eqns}
\end{equation}
By moving $a_1$ in the first equation and $a_N$ in the last equation to the right-hand-side, (\ref{eq: LN-2eqns}) can be written as a square $(N-2)\times(N-2)$ system with unknowns $a_2, \dots, a_{N-1}$ that can be computed linearly in terms of $a_1, a_N$ (under appropriate conditions of invertibility of the resulting coefficient matrix).   For the case of $N=3$, Eq.~(\ref{eq: LN-2eqns}) is one equation which we solve for $a_2$ in terms of $a_1, a_3$ to obtain
\begin{equation}
a_2 =  \frac{ a_1 + a_3 }{V_2+\omega}
\stackrel{by \ (\ref{eq: Lextansatz})}{=}
 \frac{c_1 A+ d_3 C}{V_2+\omega}
\label{eq: LN=3a2}
\end{equation}
for $\omega\neq -V_2$.

  Continuing with the $N=3$ case, we use  (\ref{eq: Lextansatz}) and (\ref{eq: LN=3a2})  to rewrite the two  remaining equations in (\ref{eq: DLS}) with $n=1,3$ as
\begin{equation}
P
\left[
\begin{array}{c}
a_0\\
a_1\\
a_2\\
a_{3}\\
a_4
\end{array}
\right]
=
PQ
\left[
\begin{array}{c}
A\\
C
\end{array}
\right]
=0
\mbox{\quad for \quad}
P=
\left[
\begin{array}{ccccc}
1 & -(V_1+\omega) & 1 & 0 & 0\\
0 & 0 & 1  & -(V_3+\omega) & 1
\end{array}
\right]
\mbox{\quad and \quad}
Q=
\left[
\begin{array}{cc}
c_0 & 0\\
c_1 & 0\\
 \frac{c_1}{V_2+\omega}&  \frac{d_3}{V_2+\omega}\\
0 & d_3 \\
0 & d_4
\end{array}
\right].
\label{eq: N=3PQlin}
\end{equation}
Again the solvability condition is $det(PQ)=0$ so that $z$ is a root of a polynomial with coefficients in terms of $V_n$.  For each $z$, one computes $A, C$ in the nullspace of $PQ$ from (\ref{eq: N=3PQlin}) so that by (\ref{eq: omegaz}), (\ref{eq: Lextansatz}) and (\ref{eq: LN=3a2}), an eigenvalue $\omega$ and
the corresponding eigenvector $a_n$ are obtained.  The $N>3$ case is similar in that $P$ and $Q$ are determined using (\ref{eq: Lextansatz}) and the above described process of using (\ref{eq: LN-2eqns}) to obtain expressions for $a_2, \dots, a_{N-1}$ in terms of $A, C$.

Due to the condition obtained in (\ref{eq: LN=3a2}), if there exists an eigenvalue  $\omega=-V_2$ then it will not be found by the above process.  In the case that such an eigenvalue exists the corresponding eigenvector is found by using (\ref{eq: Lextansatz}) to rewrite the three $n=1, 2, 3$ equations in (\ref{eq: DLS}) as the $2\times 2$ system
\begin{equation}
\left[
\begin{array}{cc}
 (V_1-V_2)c_1-c_0 & (V_2-V_3) d_3 +d_4\\
c_1 & d_3
\end{array}
\right]
\left[
\begin{array}{c}
A\\
C
\end{array}
\right]=0
\end{equation}
for $z$ such that $z+\frac{1}{z}=-V_2$.  Then  $a_n$ is computed by (\ref{eq: Lextansatz}) and $a_2=(V_1-V_2)a_1 - a_0 = (V_3-V_2)a_3-a_4$.

\subsection{Numerical results}
\label{sec: Lnum}

First, we focus on the $\cP \cT$-symmetric case where the dimer has linear potential $V_1=-V_2=i\gamma$
and the trimer potential is $V_1=-V_3=i\gamma$, while $V_2=0$.
Fig. \ref{lineigvals} shows agreement between the eigenvalues $\omega$ computed numerically
 directly from (\ref{eq: DLS}), through an eigenvalue solver,
 as compared to the semi-analytic calculation involving the identification
 of the roots of an equation of the form of Eq.~(\ref{eq: lin2V1=-V2polyn}), as
outlined in the above theoretical analysis section.
Fig. \ref{lineigvects} shows the accuracy of the semi-analytic calculation
in predicting the modulus profile of the eigenvectors $|a_n|$.

The spectra in Fig. \ref{lineigvals} have both, as expected,
a continuum component filling densely the  interval $[-2,2]$
on the real axis corresponding to the propagation of linear waves.
For $\gamma>\gamma_c$ the solutions become unstable~\cite{footnote2}; this
is the so-called $\cP \cT$-phase transition of~\cite{R1,R2,R21}.
In the case of the dimer,
the principal
unstable eigenfrequency pair has zero real part and is purely imaginary.
In the case of the trimer, on the other hand,
 there is some oscillatory behaviour superimposed
to the exponential growth of perturbations.
The exponential localization of the associated eigenvectors is
associated with the localized nature of the embedded defect structure
and is showcased in Fig.~\ref{lineigvects}.

Finally, as regards the linear eigenvalue problem,
we consider the Hamiltonian case where the dimer has a linear potential
$V_1=V_0(1+\delta)$, $V_2=V_0(1-\delta)$ and the trimer potential is
$V_1=V_0(1+\delta)$, $V_2=0$, $V_3=V_0(1-\delta)$.
Fig.~\ref{linHameigvalsvects} shows  agreement between the numerically
computed and semi-analytically calculated eigenvalues and the
(modulus of the localized) eigenvectors.  A fundamental
difference here concerns the Hermitian nature of the relevant (matrix)
operator which excludes the possibility of imaginary eigenfrequencies.
 Nevertheless, there exists within the spectrum a
real defect eigenvalue. For $\delta>0$, this
frequency
decreases as $\delta$ increases and the corresponding eigenvector profiles
for $\delta>0$ are typically similar to
what is seen in Fig. \ref{linHameigvalsvects}.

\section{Nonlinear Case}
\label{sec: N}

Let us now consider the nonlinear Schr\"odinger equation \cite{Kevrekidis}
\begin{equation}
\label{eq: TDDNS}
i\dot\Phi_n-V_n\Phi_n+\Phi_{n+1}+\Phi_{n-1} = \alpha_n |\Phi_n|^2 \Phi_n
\end{equation}
for $\alpha_n\in\mathds{R}$ zero everywhere except $1\leq n\leq N$.
Seeking, as is customary, stationary solutions of the form $ \Phi_n(t) = \psi_n e^{-i\omega t}$,
we find that they should satisfy
\begin{equation}
\omega \psi_n -V_n\psi_n +\psi_{n+1} +\psi_{n-1} =\alpha_n|\psi_n|^2 \psi_n.
\label{eq: Nstatpsieqn}
\end{equation}
We consider the dynamics of small perturbations defined by
\begin{equation}
\Phi_n(t) = (\psi_n + \epsilon \phi_n) e^{-i\omega t}; \quad
\phi_n\equiv\left( a_ne^{i\nu t}+b_n e^{-i\nu^* t} \right)
\label{eq: phipsiab}
\end{equation}
for $\omega \in\mathds{R}$ and $a_n, b_n,\nu \in\mathds{C}$ and with $\psi_n \in\mathds{C}$.
In order to investigate the stability, we examine the resulting order-$\epsilon$ equations,
amounting to the spectral or linear stability analysis equations
\begin{equation}
i \dot \phi_n = (V_n-\omega)\phi_n
-\phi_{n+1} - \phi_{n-1} + \alpha_n \left(2|\psi_n|^2 \phi_n
+\psi_n^2 \phi_n^*\right).
\label{linear}
\end{equation}
Note that $\phi_n$ is complex. It is thus recognized that
problem (\ref{linear}) is similar to (\ref{eq: TDDLS}), the main difference
being that now  $\phi_n$ is coupled to $\phi_n^*$.
Hence, given a zeroth-order solution $\psi_n$,
we can apply a similar approach as the one developed in
section~\ref{sec: L}, in order to determine its linearization spectrum.

For later reference, it is convenient to reformulate the problem in matrix form
for $a_n, b_n$ which are obtained from  (\ref{eq: TDDNS}) by equating
coefficients of $e^{i (\nu-\omega)t}, e^{-i(\nu^*+\omega)t}$, yielding
 \begin{equation}
 \nu \left[  \begin{array}{c} a_n\\ b_n^* \end{array}  \right]
 =
F
\left[  \begin{array}{c} a_n\\ b_n^* \end{array}  \right]
\label{eq: Oepsmatrix}
\end{equation}
for $F= \left[ \begin{array}{cc} F_1 & F_2\\ F_3 & F_4 \end{array} \right]$ and
\begin{eqnarray}
\begin{array}{cclcccl}
F_1&=& {\rm diag}(\omega-V_n-2\alpha_n|\psi_n|^2)+G &\quad&F_2&=& -{\rm diag}(\alpha_n\psi_n^2)\\
F_3&=& {\rm diag}(\alpha_n(\psi_n^*)^2) &\quad& F_4&=&-{\rm diag}(\omega-V_n^*-2\alpha_n|\psi_n|^2)-G
\end{array}
\label{eq: FwrtValphapsi}
\end{eqnarray}
where $G$ is a sparse matrix with ones on the first super- and sub-diagonal.

\subsection{Theoretical Analysis}
\label{sec: Nthy}

Similar to the linear case we begin with the linearization
problem ansatz
\begin{equation}
a_n=
\left\{
\begin{array}{cc}
Ae^{i\kappa n} + Be^{-i\kappa n} &n<1\\
Ce^{i\kappa n} + De^{-i\kappa n} & n>N
\end{array}
\right. ,
\qquad
\label{eq: Nansatz}
b_n^*=
\left\{
\begin{array}{cc}
A'e^{i\kappa' n} + B'e^{-i\kappa' n} &n<1\\
C'e^{i\kappa' n} + D'e^{-i\kappa' n} & n>N
\end{array}
\right.
\end{equation}
for $A, C, B, D,  A', C',  B', D'\in\mathds{C}$ and $\kappa, \kappa' \in\mathds{C}$ satisfying dispersion relations
\begin{equation}
\nu-\omega = 2\cos(\kappa) \qquad    -(\nu +\omega) = 2\cos(\kappa').
\label{eq: Ndisprel}
\end{equation}
Imposing (\ref{eq: Nansatz}) and (\ref{eq: Ndisprel}) is enough to satisfy all equations in (\ref{eq: Oepsmatrix}) except those with $n=0, 1, \dots, N, N+1$.  Applying (\ref{eq: Nansatz}) to (\ref{eq: Oepsmatrix}) also shows that formulae (\ref{eq: Nansatz}) are applicable at $n=1, N$.  Homogeneous Dirichlet boundary conditions $a_{n_0-1}=b^*_{n_0-1}=a_{m_0+1}=b^*_{m_0+1}=0$ (again for simplicity/specificity) imply that $\nu - \omega  = {a_{n_0+1}}/{a_{n_0}} = { a_{m_0-1} }/{a_{m_0}}$ and $-(\nu + \omega)= { b^*_{n_0+1}}/{b^*_{n_0}} = {b^*_{m_0-1}}/{b^*_{m_0}}$ which when combined with the ansatz (\ref{eq: Nansatz}) gives $B = -Ae^{i\kappa (2n_0-2)}$, $D = -Ce^{i\kappa (2m_0+2)}$, $B' = -A'e^{i\kappa' (2n_0-2)}$, $D' = -C'e^{i\kappa' (2m_0+2)}$.  Thus we have
 \begin{equation}
  a_n=
  \left\{
  \begin{array}{cc}
  A c_n&n\leq 1\\
  C d_n & n\geq N
  \end{array}
  \right. ,
  \qquad
  \label{eq: Nextansatz}
  b_n^*=
  \left\{
  \begin{array}{cc}
  A' c'_n &n\leq 1\\
  C' d'_n & n\geq N
  \end{array}
  \right.
  \end{equation}
  where $c_n$ is defined in terms of $z=e^{i\kappa}$ as before and $c'_n=(z')^{ n} - (z')^{ 2n_0-2-n}$, $d'_n = (z')^{ n} - (z')^{2m_0+2-n}$ for $z'=e^{i \kappa'}$.

When combined with the dispersion relations (\ref{eq: Ndisprel}),
the condition (\ref{eq: Nextansatz}) is enough to satisfy all except the $2N$ equations in (\ref{eq: Oepsmatrix})   associated with $n=1, \dots, N$.  Again we separate the remaining parts of the computation into two sections.

\subsubsection{\bf \emph{Oligomer of length two}}
\label{sec: Nthy2}

For $N=2$ the remaining four equations in (\ref{eq: Oepsmatrix}) for $n=1,2$ can be written
 as $Pv=0$ for $P= \left( \begin{array}{cc} P_1 & P_2\\ P_3 & P_4 \end{array} \right)$ and
\begin{equation}
P_1=
\left[
\begin{array}{ccccc}
1 & \omega-\nu-V_1-2\alpha_1|\psi_1|^2 &             1                &0 \\
  0 &   1 & \omega-\nu-V_2-2\alpha_2|\psi_2|^2 &             1              \\
\end{array}
\right], \quad
P_2=
\left[
\begin{array}{ccccc}
0 & -\alpha_1\psi_1^2 &             0                &0 \\
  0 &   0 & -\alpha_2\psi_2^2 &             0              \\
\end{array}
\right],
\nonumber
\end{equation}
\vspace{-.25in}
\begin{equation}
\vspace{-.2in}
\label{eq: N=2nlinP}
\end{equation}
\begin{equation}
P_3=
\left[
\begin{array}{ccccc}
0 & \alpha_1\left(\psi_1^*\right)^2 &             0                &0 \\
  0 &   0 & \alpha_2\left(\psi_2^*\right)^2 &             0              \\
\end{array}
\right], \quad
P_4=
\left[
\begin{array}{ccccc}
-1 & -\omega-\nu+V_1^*+2\alpha_1|\psi_1|^2 &             -1                &0 \\
  0 &   -1 & -\omega-\nu+V_2^*+2\alpha_2|\psi_2|^2 &             -1              \\
\end{array}
\right],
\nonumber
\end{equation}
where $v=\left[ \begin{array}{c}a_n \\ b_n^* \end{array} \right]$ is a vector of length eight with each of $a_n, b_n^*$  restricted to the index $0\leq n\leq 3$.  By (\ref{eq: Nextansatz}), $v$ can be written as
\begin{equation}
v=Q
\left[
\begin{array}{c}
A\\
C\\
A'\\
C'
\end{array}
\right]
\mbox{ \quad for \quad }
Q=
\left[
\begin{array}{cccc}
c_0 & 0 & 0 & 0\\
c_1 & 0 & 0 & 0\\
0 & d_2 & 0 & 0\\
0 & d_3 & 0 & 0\\
0 & 0 & c'_0 & 0\\
0 & 0 & c'_1 & 0\\
0 & 0 & 0 & d'_2\\
0 & 0 & 0 & d'_3\\
\end{array}
\right].
\end{equation}
The solvability condition is then the pair of equations
\begin{equation}
det(PQ)=0, \qquad z+\frac{1}{z}+\omega = -\left( z'+\frac{1}{z'}+\omega\right)
\label{eq: det=0,nu=nu}
\end{equation}
where the second was obtained from the dispersion relations (\ref{eq: Ndisprel}).  Solutions of (\ref{eq: det=0,nu=nu}) where either of $z$ or $z'$ is $\pm 1$ can be disregarded since the resulting system $Pv=0$ has no non-trivial solutions.  The remaining solutions appear in quadruple sets  $(z, z')$, $(z, \frac{1}{z'})$, $(\frac{1}{z}, z')$, $(\frac{1}{z}, \frac{1}{z'})$ so that the total number of relevant solutions of (\ref{eq: det=0,nu=nu}) is four times the length of the lattice.  Thus by (\ref{eq: Ndisprel}) the calculation yields a number (counted with algebraic multiplicity) of eigenvalues $\nu$ equal to the length of the lattice.

\begin{figure}[tbp]
\begin{center}
\includegraphics[width=8cm,angle=0,clip]{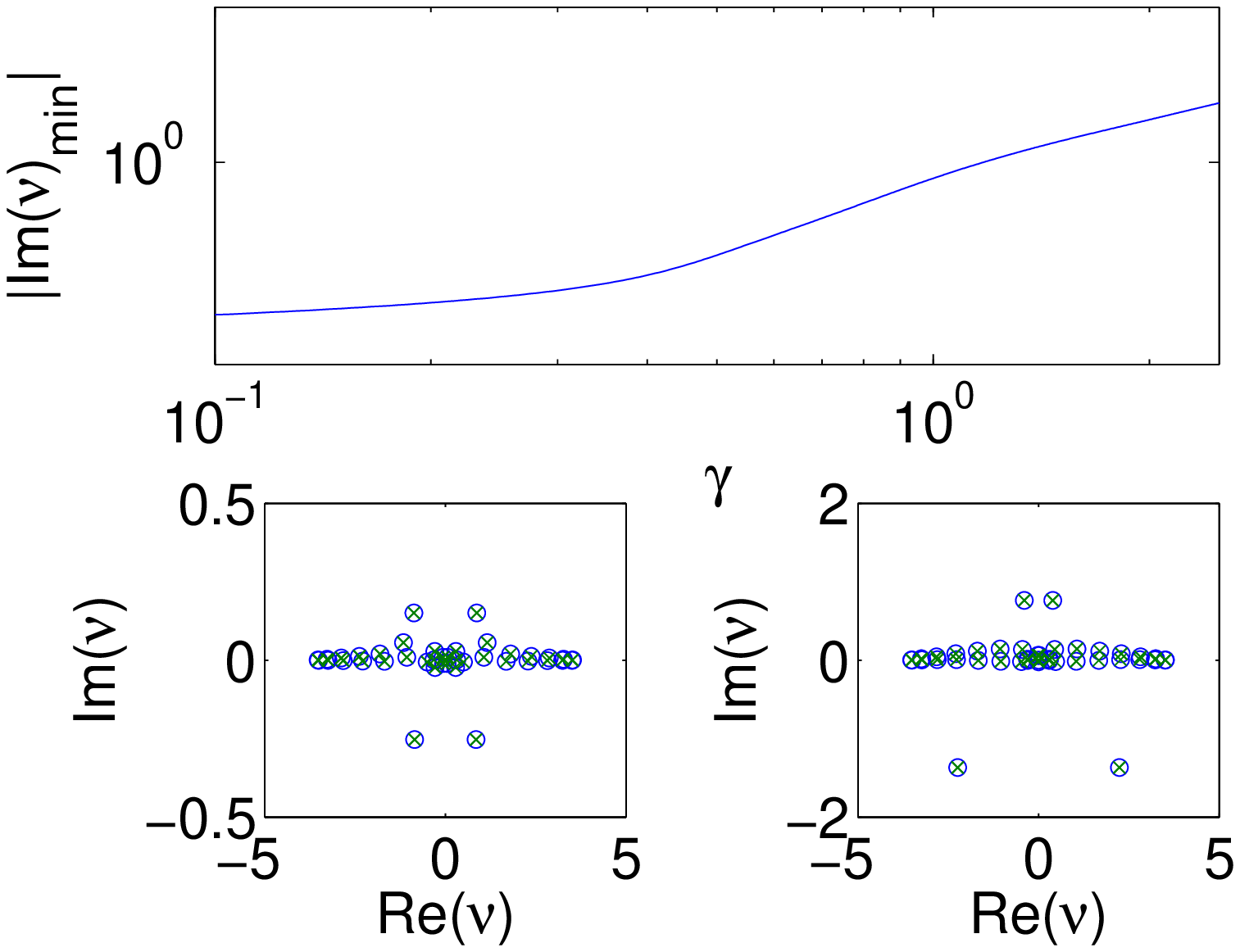}
\includegraphics[width=8cm,angle=0,clip]{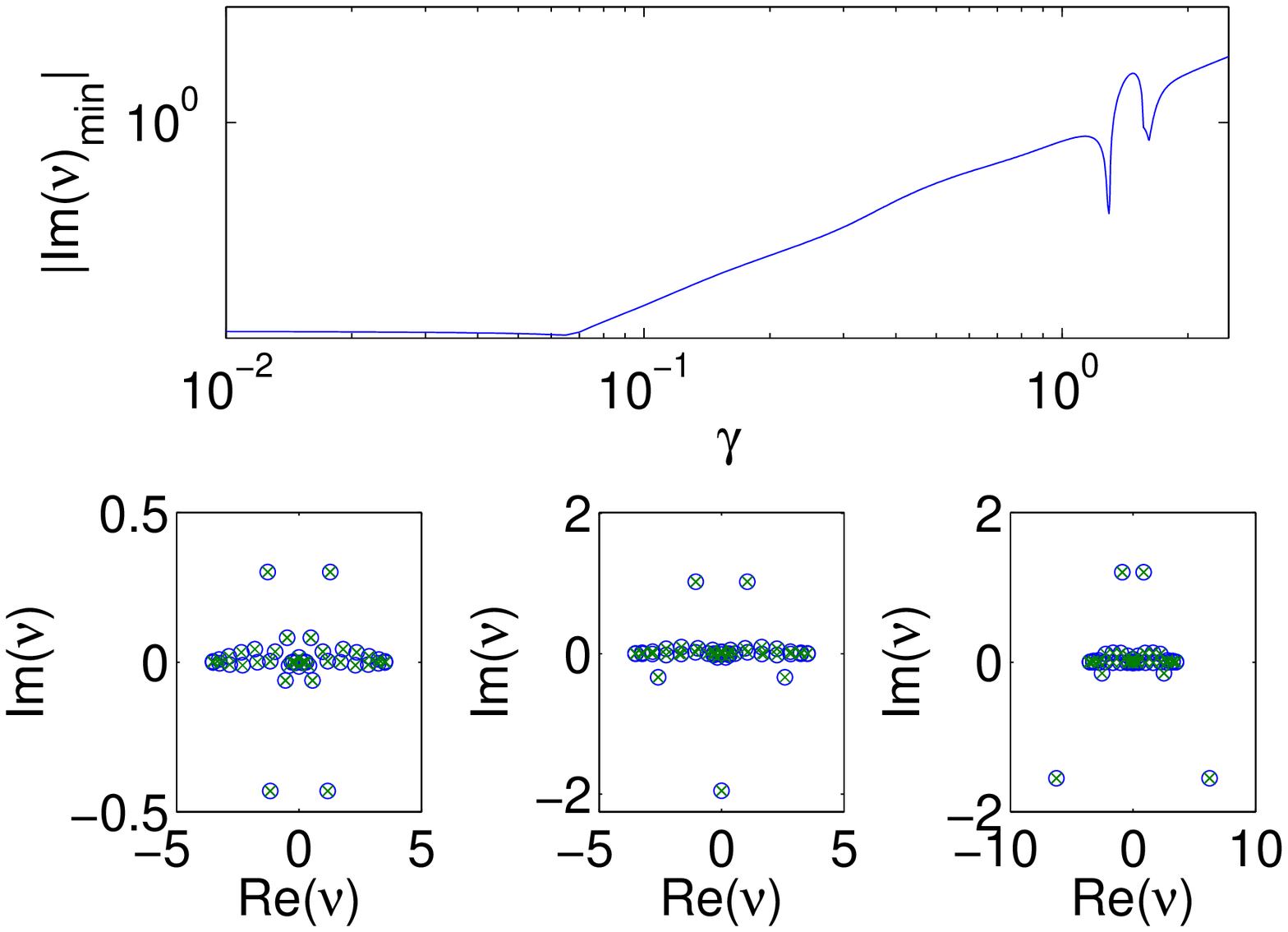}
\end{center}
\caption{
The left set of panels corresponds to the $\cP \cT$-symmetric dimer with $N=2$, $\alpha_{1,2}=1$ and $V_1=i\gamma=-V_2$ while the right set of panels to the case of the trimer with $N=3$, $\alpha_{1,2,3}=1$ and $V_1=i\gamma=-V_3, V_2=0$.  Each set contains a (top) plot of linear stability eigenfrequencies of the
stationary solution corresponding to $T=0.7$ and $k_0=2.5$ as computed on a lattice of length $20$.
Plots of the eigenvalue $\nu$ indicating
agreement between numeric (circles) and exact (x's) results are shown for the
dimer with $\gamma=.5$ (bottom left) and $\gamma=1.5$ (bottom right), and for
the trimer with $\gamma=.5$ (bottom left) and $\gamma=1.5$ (bottom middle)
and $\gamma=1.75$ (bottom right).
For the trimer, when $\gamma$ is small the dominant pair of (unstable)
eigenfrequencies of negative imaginary part increases in magnitude
as $\gamma$ increases until $\gamma\approx 1.15$.  At this point the
complex pair recedes and a single dominant purely imaginary
emerges at $\gamma\approx 1.3$.   This single eigenfrequency
increases in magnitude until $\gamma\approx 1.49$ when it begins to
decrease.  At $\gamma\approx 1.62$, a new dominant complex pair
increases in magnitude and continues to increase as $\gamma$ increases.
}
\label{nlineigvals}
\end{figure}

\begin{figure}[tbp]
\begin{center}
\includegraphics[width=8cm,angle=0,clip]{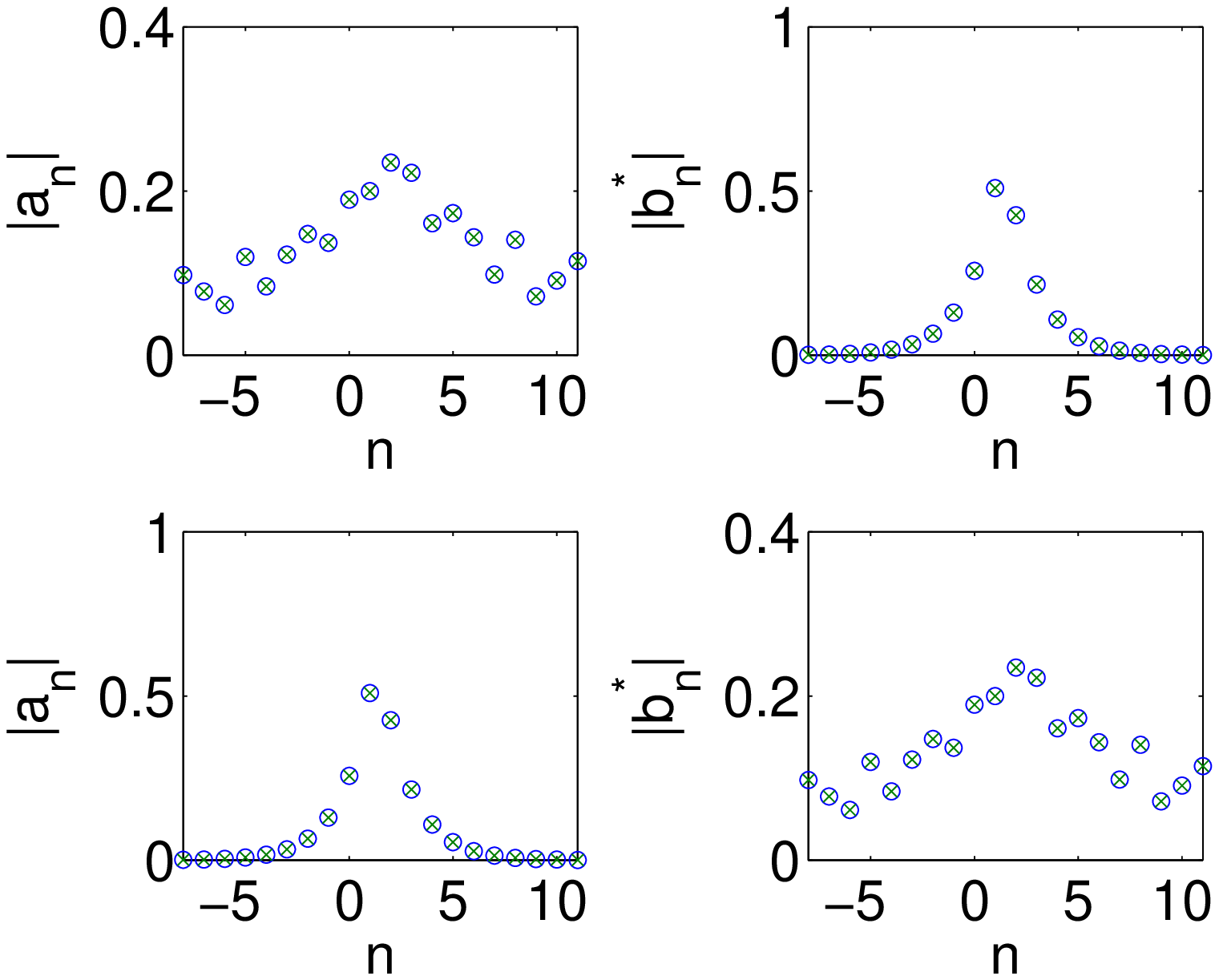}
\includegraphics[width=8cm,angle=0,clip]{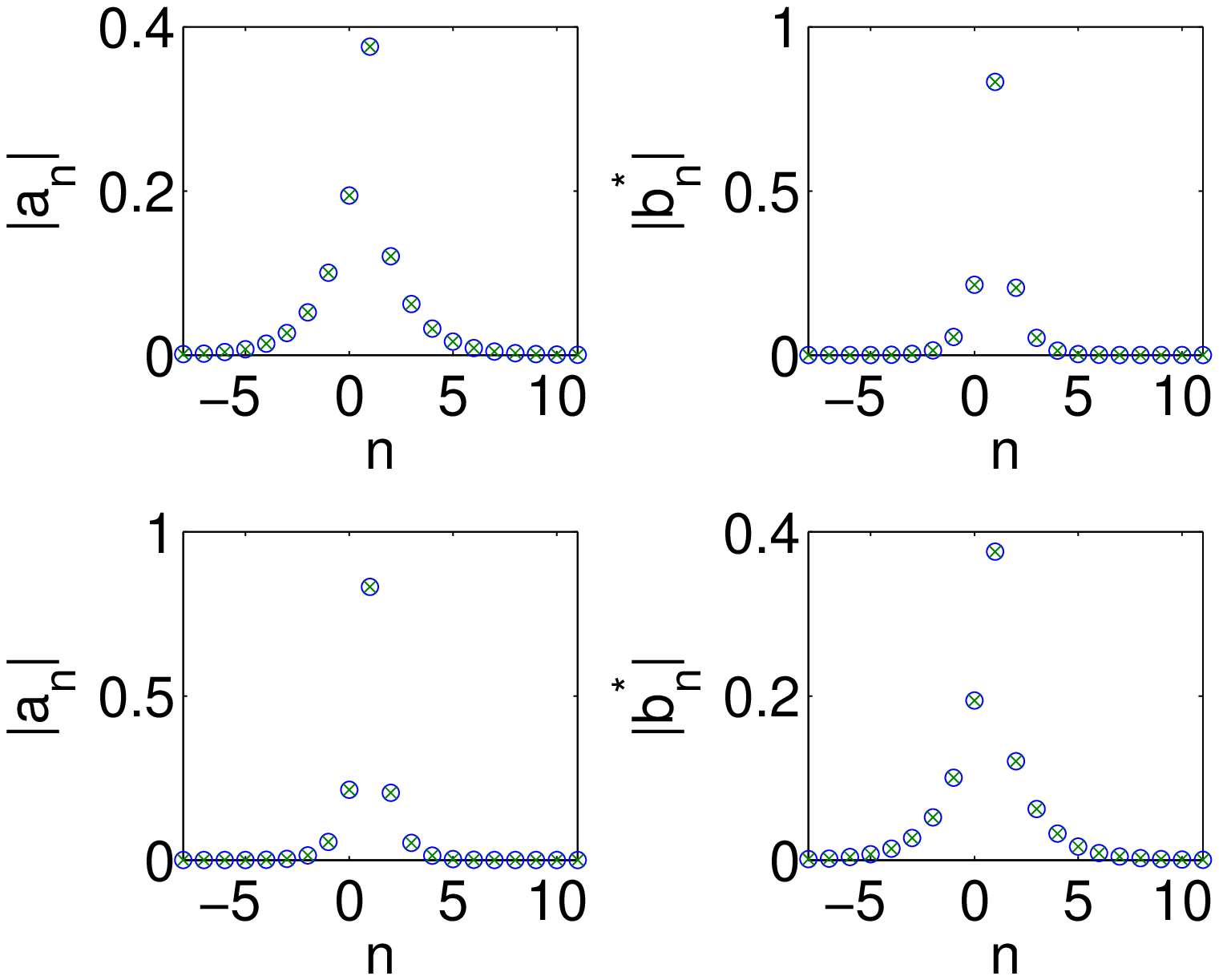}
\end{center}
\caption{
Profiles of the moduli of unstable eigenvectors $|a_n|$, $|b^*_n|$ are shown for the nonlinear dimer.
The left four plots show eigenvectors for $\gamma=.5$  and the right four plots for $\gamma=1.5$.  These eigenvectors correspond to eigenvalues with negative imaginary part which are seen in Fig. \ref{nlineigvals}.  Again, numerical computations are shown to agree with the exact (x's) results and the length of the lattice is $20$.
Notice the localization in both components of the eigenvector in the
case of the $\cP \cT$-symmetry breaking case of $\gamma>1$.
}
\label{nlindimereigvects}
\end{figure}

\begin{figure}[tbp]
\begin{center}
\includegraphics[width=8cm,angle=0,clip]{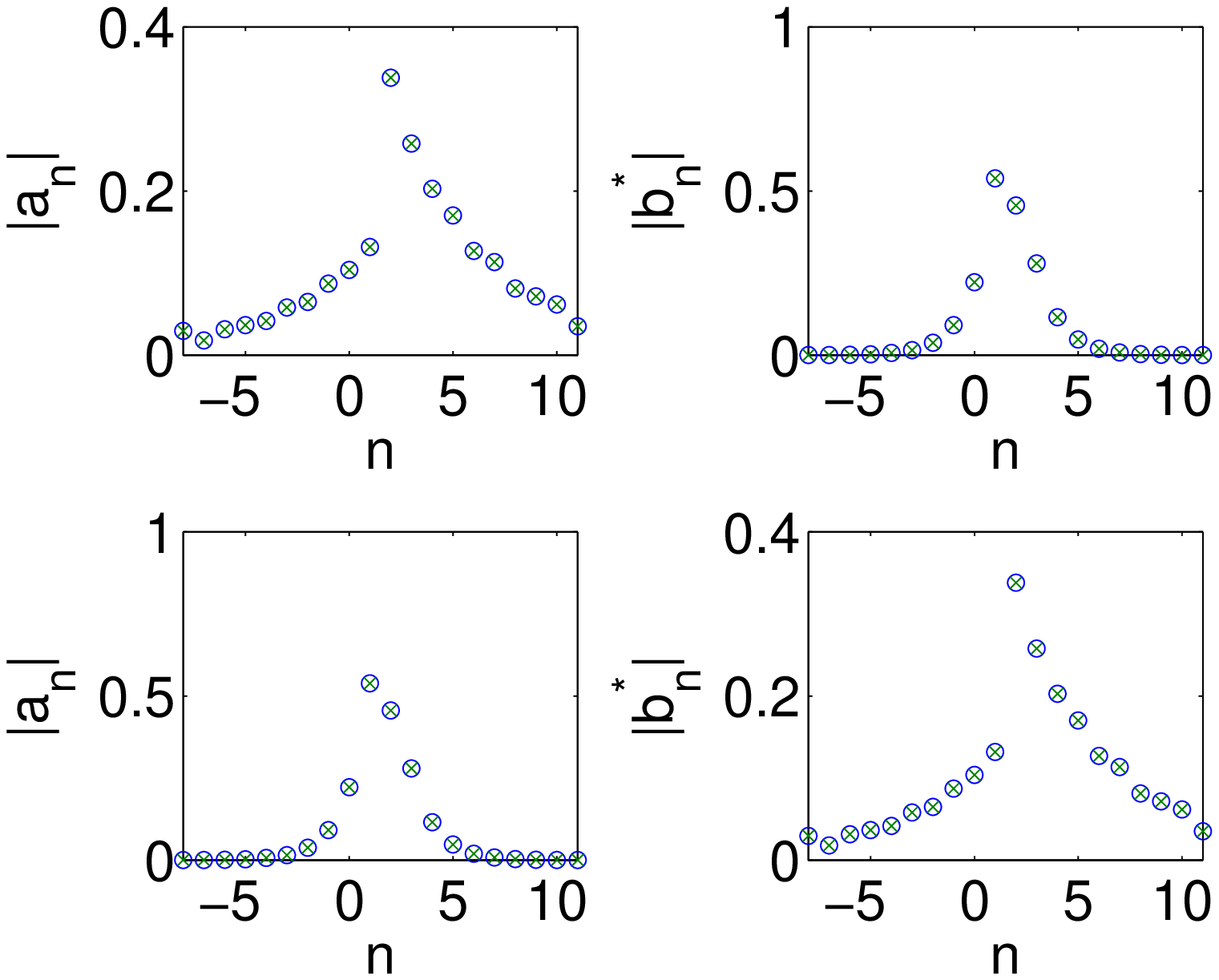}
\includegraphics[width=8cm,angle=0,clip]{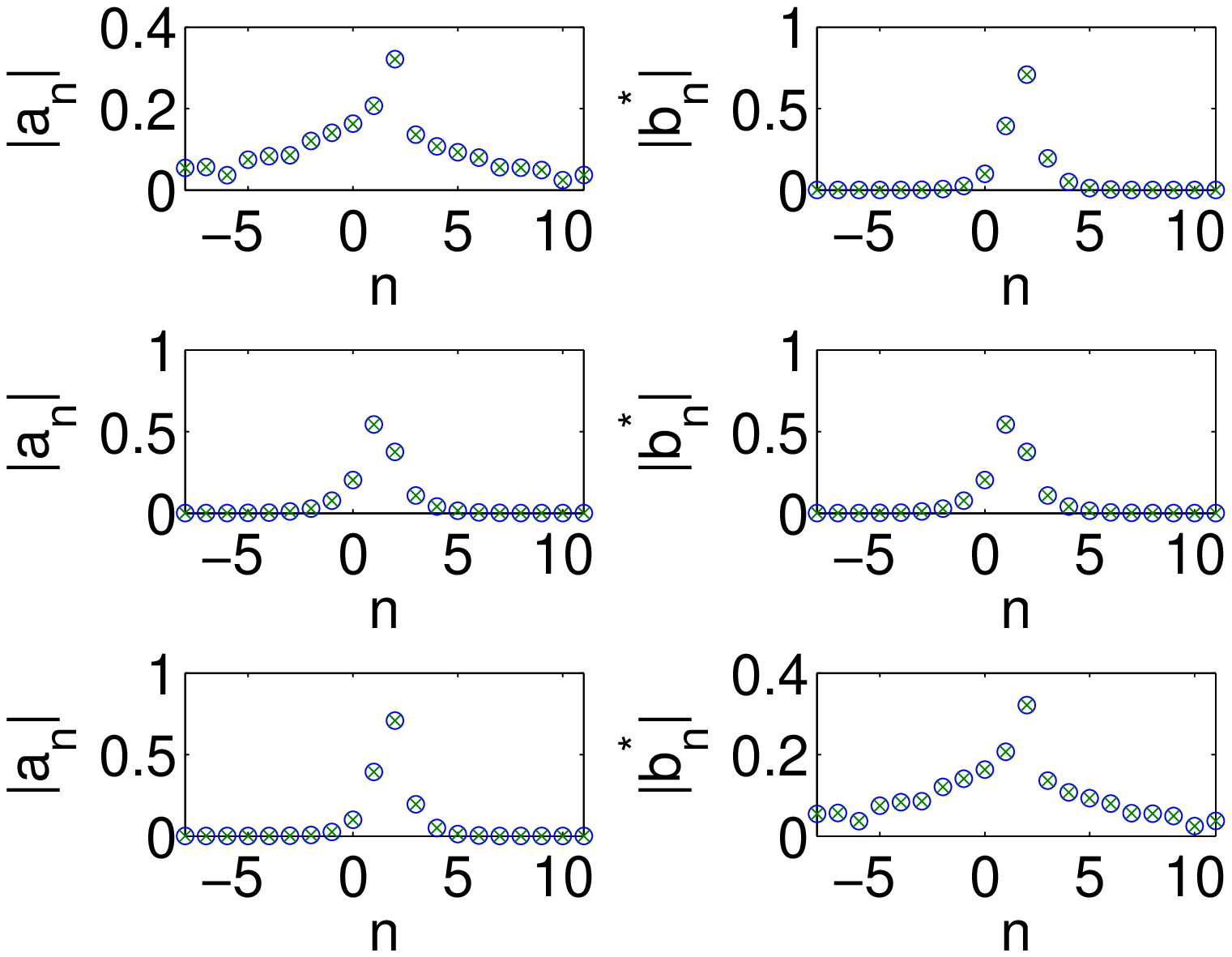}
\end{center}
\caption{
Profiles of the moduli of unstable eigenvectors $|a_n|$, $|b^*_n|$ are shown for the nonlinear trimer with lattice length $20$.
The left four plots show eigenvectors for $\gamma=.5$ and the right six plots for $\gamma=1.5$.  These eigenvectors correspond to eigenvalues with negative imaginary part which are seen in Fig. \ref{nlineigvals}.
The agreement is similar to those of the earlier figures.
}
\label{nlintrimereigvects}
\end{figure}

\subsubsection{\bf \emph{Oligomer of length three or greater}}
\label{sec: Nthy>2}

Similarly to the strategy for the linear case, for $N>2$ we begin by examining the $2(N-2)$ equations in (\ref{eq: Oepsmatrix}) with $n=2, \dots, N-1$.  Using (\ref{eq: Nextansatz}) to rewrite  the four quantities $a_1, a_N, b^*_1, b^*_N$ one obtains a non-homogeneous square system where the $2(N-2)$ variables $a_2, \dots, a_{N-1}, b^*_2, \dots, b^*_{N-1}$ can be computed as linear combinations of $A, C, A', C'$, under appropriate conditions of invertibility of the coefficient matrix.

In the case of $N=3$ the $2(N-2)=2$ equations can be written as
\begin{equation}
\left[
\begin{array}{cccccc}
1 & \omega-\nu-V_2-2\alpha_2|\psi_2|^2 & 1 & 0 & -\alpha_2 \psi_2^2 & 0\\
0 & \alpha_2\left( \psi_2^* \right)^2 & 0 & -1 & -\omega-\nu+V^*_2+2\alpha_2 |\psi_2|^2 & -1
\end{array}
\right]
\left[
\begin{array}{c}
a_n\\
b^*_n
\end{array}
\right]
=0\label{eq: N=3extrastep}
\end{equation}
where the column vector is length six with $1\leq n\leq 3$.  We move $a_1, a_3, b^*_1, b^*_3$ to the right-hand-side of (\ref{eq: N=3extrastep}) and use (\ref{eq: Nextansatz}) to obtain
\begin{eqnarray}
\left[
 \begin{array}{c}
 a_2\\
 b^*_2
 \end{array}
 \right]
 =
 M
\left[
\begin{array}{c}
A\\
C\\
A'\\
C'
\end{array}
\right]
\mbox{ \quad for \quad}
M=
 \left[
 \begin{array}{cc}
 \omega-\nu-V_2-2\alpha_2|\psi_2|^2 & -\alpha_2 \psi_2^2 \\
 \alpha_2 \left( \psi^*_2\right)^2 & -\omega-\nu+V^*_2+2\alpha_2|\psi_2|^2
 \end{array}
 \right]^{-1}
 \left[
\begin{array}{cccc}
-c_{1} & -d_{3} & 0 & 0\\
0 & 0 & c'_{1} & d'_{3}
\end{array}
\right].
\label{eq: nlina2b2M}
\end{eqnarray}
Of course, (\ref{eq: nlina2b2M}) is the nonlinear analogue of (\ref{eq: LN=3a2}) and the values of $z$ which are found by the remaining parts of the computation are such that the inverse matrix in (\ref{eq: nlina2b2M}) exists.  The four remaining equations in (\ref{eq: Oepsmatrix}) with $n=1,3$ can now be written as $Pv=0$ for
$P = \left( \begin{array}{cc} P_1 & P_2\\ P_3 & P_4 \end{array} \right)$ with
\begin{equation}
P_1=
\left[
\begin{array}{ccccc}
1 & \omega-\nu-V_1-2\alpha_1|\psi_1|^2 &             1                &0&0 \\
0  & 0 &   1 & \omega-\nu-V_3-2\alpha_3|\psi_3|^2 &             1              \\
\end{array}
\right],
\quad
P_2=
\left[
\begin{array}{ccccc}
0 & -\alpha_1\psi_1^2 &             0                &0&0 \\
  0& 0 &   0 & -\alpha_3\psi_3^2 &             0              \\
\end{array}
\right] \nonumber
\end{equation}
\vspace{-.25in}
\begin{equation}
\vspace{-.2in}
\label{eq: N=3nlinP}
\end{equation}
\begin{equation}
P_3=
\left[
\begin{array}{ccccc}
0 & \alpha_1\left(\psi_1^*\right)^2 &             0                &0&0 \\
  0& 0&   0 & \alpha_3\left(\psi_3^*\right)^2 &             0              \\
\end{array}
\right], \quad
P_4=
\left[
\begin{array}{ccccc}
-1 & -\omega-\nu+V_1^*+2\alpha_1|\psi_1|^2 &             -1                &0&0 \\
  0& 0&   -1 & -\omega-\nu+V_3^*+2\alpha_3|\psi_3|^2 &             -1              \\
\end{array}
\right] \nonumber
\end{equation}
and where $v=\left[\begin{array}{c}a_n\\ b^*_n\end{array}\right]$ is length ten with $0\leq n\leq 4$.  By (\ref{eq: Nextansatz}) and (\ref{eq: nlina2b2M}) we write
\begin{equation}
v=Q
\left[
\begin{array}{c}
A\\
C\\
A'\\
C'
\end{array}
\right]
\mbox{ \quad for \quad }
Q=
\left[
\begin{array}{cccc}
c_0 & 0 & 0 & 0\\
c_1 & 0 & 0 & 0\\
M_{11}&M_{12}&M_{13}&M_{14}\\
0 & d_3 & 0 & 0\\
0 & d_4 & 0 & 0\\
0 & 0 & c'_0 & 0\\
0 & 0 & c'_1 & 0\\
M_{21}&M_{22}&M_{23}&M_{24}\\
0 & 0 & 0 & d'_3\\
0 & 0 & 0 & d'_4\\
\end{array}
\right].
\label{eq: N=3nlinQ}
\end{equation}
The solvability condition is the same as (\ref{eq:  det=0,nu=nu}) but with $P$ and $Q$ as defined in (\ref{eq: N=3nlinP}) and (\ref{eq: N=3nlinQ}), respectively.  Again, solutions with either $z$ or $z'$ equal to $\pm 1$ are not relevant.

\begin{figure}[tbp]
\begin{center}
\includegraphics[width=8cm,angle=0,clip]{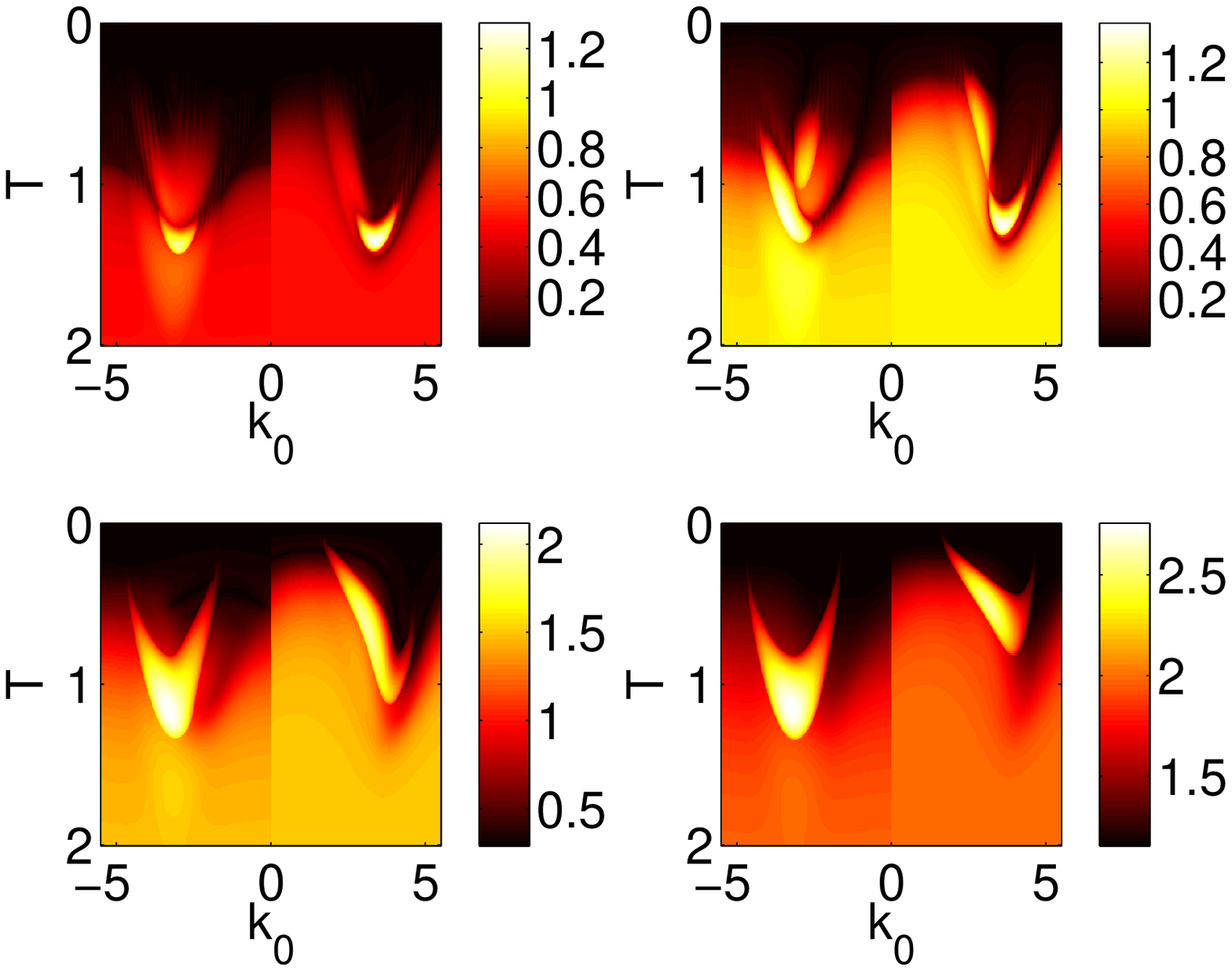}
\includegraphics[width=8cm,angle=0,clip]{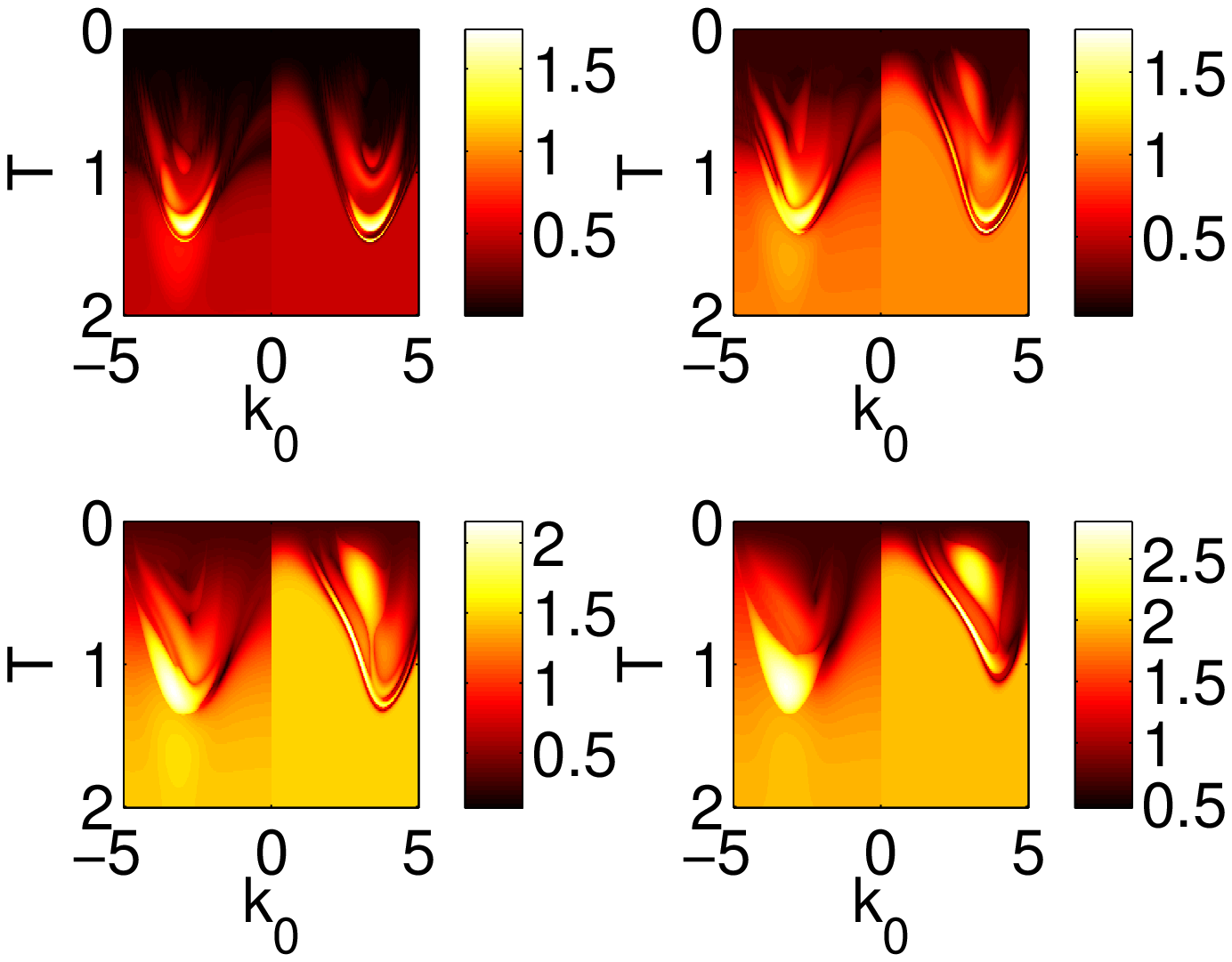}
\end{center}
\caption{
The left set of panels correspond to the $\cP \cT$-symmetric dimer with $N=2$, $\alpha_{1,2}=1$ and $V_1=i\gamma=-V_2$ while the right set of panels to the case of the trimer with $N=3$, $\alpha_{1,2,3}=1$ and $V_1=i\gamma=-V_3, V_2=0$.  Each set contains contour plots of extremal stability eigenfrequencies $|Im(\nu)_{min}|$ for extended solutions as in (\ref{eq: FormOfPsi_n}) on a lattice of length $20$ with $\gamma=.5$ (top left), $\gamma=1$ (top right), $\gamma=1.5$ (bottom left) and $\gamma=2$ (bottom right).
}
\label{absminimnuPTSymm}
\end{figure}

\begin{figure}[tbp]
\begin{center}
\includegraphics[width=8cm,angle=0,clip]{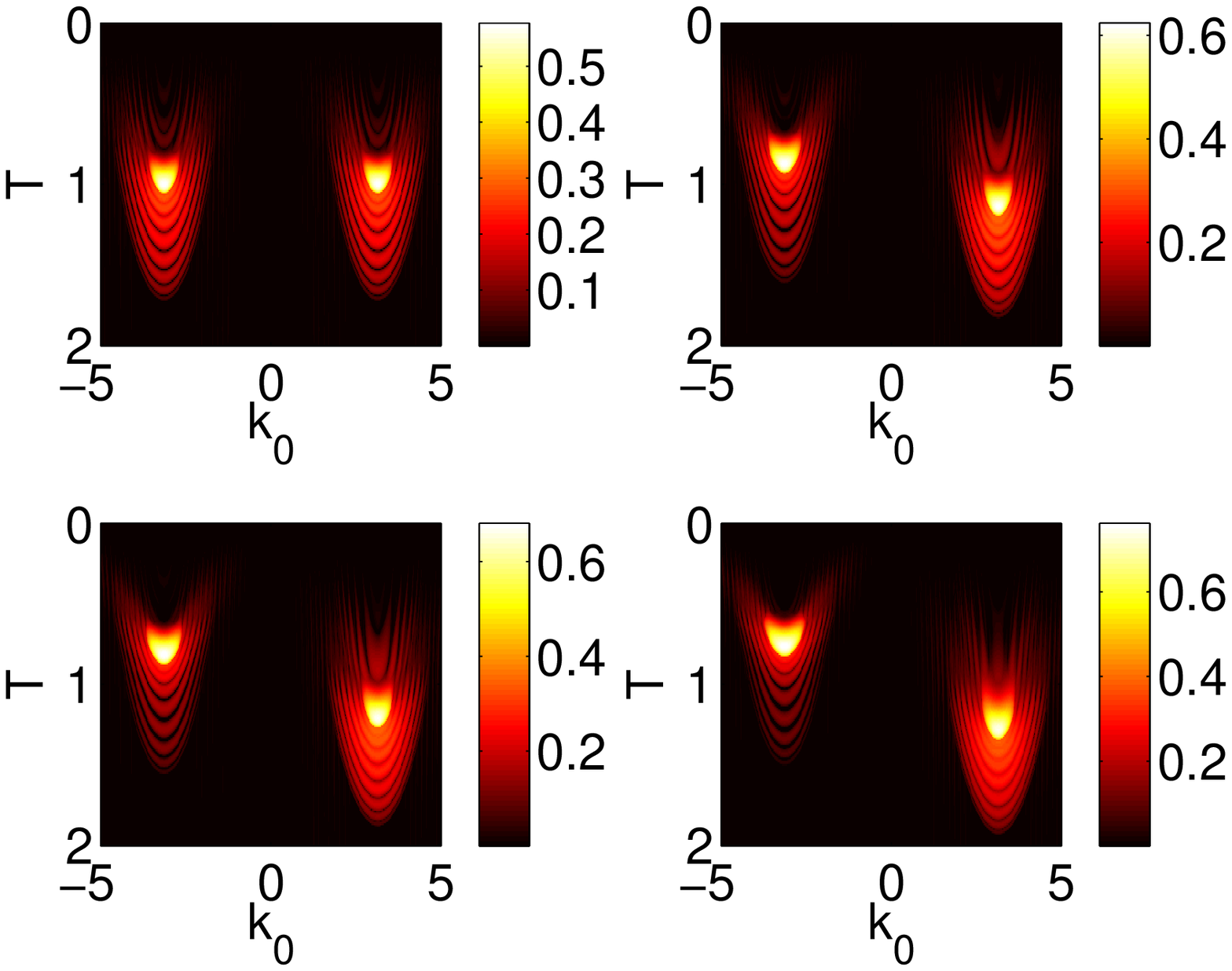}
\includegraphics[width=8cm,angle=0,clip]{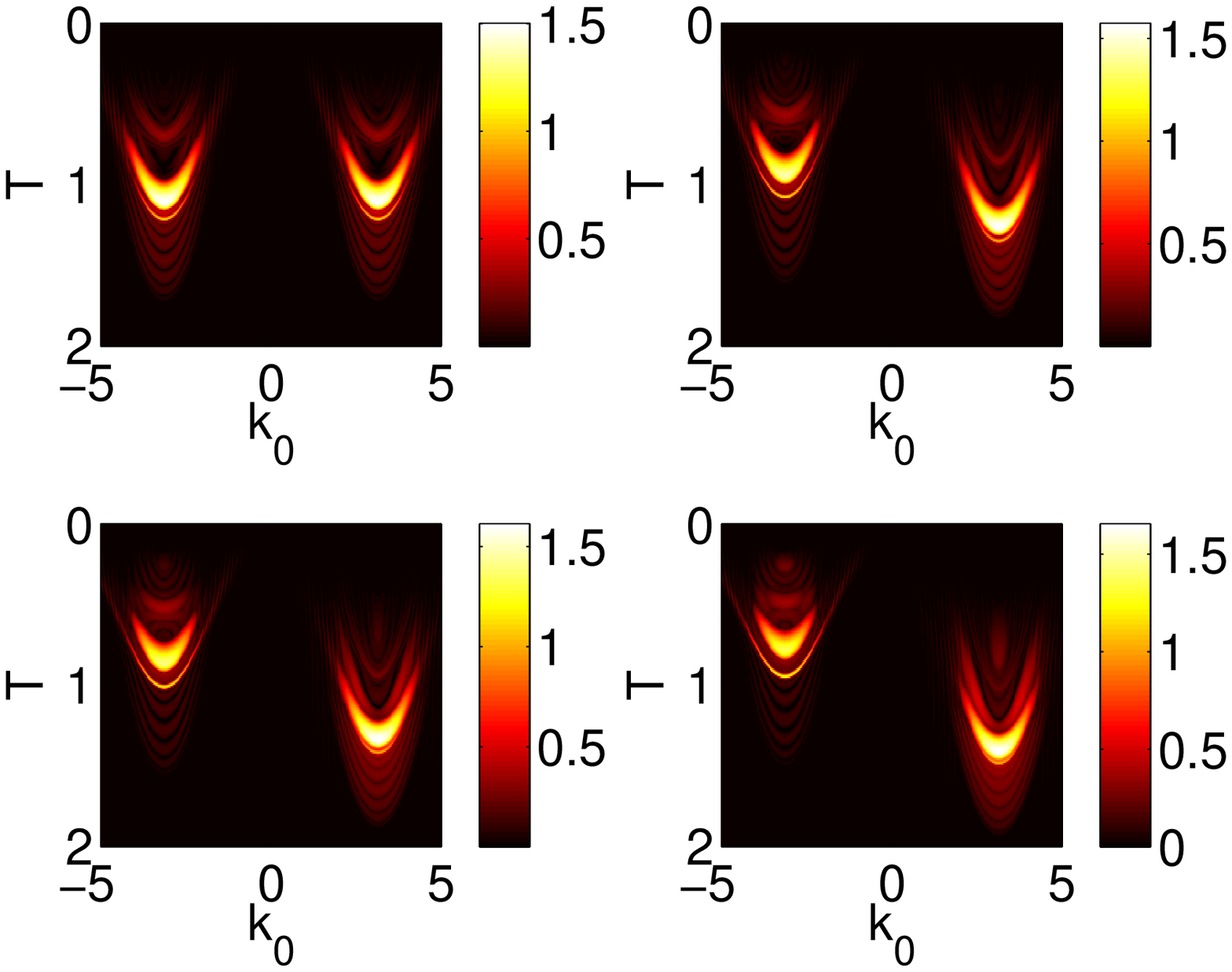}
\end{center}
\caption{
The left set of panels correspond to the Hamiltonian dimer with $N=2$, $\alpha_{1,2}=1$ and $V_1=1+\delta, V_2=1-\delta$ while the right set of panels to the case of the trimer with $N=3$, $\alpha_{1,2,3}=1$ and $V_1=1+\delta, V_2=0, V_3=1-\delta$.  Each set contains contour plots of extremal stability eigenfrequencies $|Im(\nu)_{min}|$ for extended solutions as in (\ref{eq: FormOfPsi_n}) on a lattice of length $20$ with $\delta=0$ (top left), $\delta=0.5$ (top right), $\delta=0.75$ (bottom left) and $\delta=1$ (bottom right).
}
\label{absminimnuHamilt}
\end{figure}

\subsection{Numerical results}
\label{sec: Nnum}

As mentioned in the Introduction, one of the motivations of the present
work is to study the stability of scattering solutions. In particular,
we focus here on of the class of plane wave solutions
of (\ref{eq: Nstatpsieqn}) of the form
\begin{equation}  \psi_n=\left\{
\begin{array}{ll}
R_0e^{ik_0n}+Re^{-ik_0n} & n\leq 1\\
Te^{ik_0n} & n\geq N
\end{array}
\right.
\label{eq: FormOfPsi_n}  \end{equation}
with $R_0, R, T\in\mathds{C}$ representing the incident, reflected and transmitted amplitudes, respectively
and $k_0 \ge 0$ is the wavenumber such that $\omega=-2\cos(k_0)$.
For each given values of $k_0, T$ one can compute $\psi_n$ by repeated application of the
backwards transfer map \cite{Delyon86,Wan90,Li96,Hennig99}
\begin{equation}
\psi_{n-1}=-\psi_{n+1}+(V_n-\omega+\alpha_n |\psi_n|^2)\psi_n
\label{eq: BXFerMap}
\end{equation}
which is a rearrangement of (\ref{eq: Nstatpsieqn}). For short oligomers, like the
ones we are dealing with here, the $\psi_n$ can be evaluated analytically
\cite{lepri,ourselves}.

Fig.~\ref{nlineigvals} shows an example of the the eigenvalues $\nu$ computed with
the method discussed in the previous section. As a check, we also report eigenvalues
computed by numerical diagonalization of the matrix $F$ in (\ref{eq: Oepsmatrix})
evaluated by substituting the computed $ \psi_n$ in (\ref{eq:  FwrtValphapsi}).
The corresponding eigenvectors for the case of the dimer
and the trimer are shown, respectively, in Figs.~\ref{nlindimereigvects}
and \ref{nlintrimereigvects}.

It is clear that past the critical point
of the $\cP \cT$ phase transition, the relevant eigenvectors become
highly localized (contrary to what is the case with the more spatially
extended eigenvectors before the transition). These eigenvectors
are responsible for the rapid growth of the norm density at the
gain node observed previously ~\cite{ourselves}.
The analysis also confirms the finding
from our previous work \cite{ourselves} that higher values of $\gamma$ and
$T$ correspond to more unstable solutions.  We partially
capture this phenomenon in Fig. \ref{absminimnuPTSymm},
which contains a systematic two-parameter diagram of the dependence of
the growth rate of the corresponding most unstable eigenstates.
In addition to the stronger instability for higher $T$ observed
in the figure (and also for higher $\gamma$ in Fig.~\ref{nlineigvals}),
we observe the asymmetry of the relevant growth rate for transmission
in the direction of positive vs. negative $k_0$ (as in
\cite{lepri,ourselves} we have adopted the convention that $-k_0$ labels a
right-incoming solution with wavenumber $k_0$).  Among these
two directions, it is intuitively clear and numerically confirmed that
larger growth rates are incurred for waves that first encounter
the gain site.   { For a fixed $\gamma$, the transition from right-incoming ($-k_0<0$) to left-incoming ($k_0>0$) waves may be achieved computationally by allowing the wave number to stay positive and instead flipping the potential $V_n$ from left to right.  For the $\cP\cT$ case this amounts to a continuous change in $k_0$ and a discontinuous change in the fixed $\gamma$ value from negative to positive.  This explains the discontinuity apparent in Fig. \ref{absminimnuPTSymm}.
}

Finally, we touch upon the Hamiltonian case namely the case of real-valued $V_n$
for the nonlinear problem \cite{lepri} [once again we use a dimer with
$V_1=V_0(1+\delta)$, $V_2=V_0(1-\delta)$ and a trimer with
$V_1=V_0(1+\delta)$, $V_2=0$, $V_3=V_0(1-\delta)$].  This is captured in
Fig. \ref{absminimnuHamilt}.   In the Hamiltonian case the instability for the dimer associated with the particular values $k_0=2.5$ and $T=0.7$, for example, increases as $\delta$ increases from zero to $\delta\approx 1.35$ then the instability decreases for increasing $\delta$.  Additional oscillations appear in Figure 8 as an artifact of the short lattice length which we used to obtain the higher resolution with respect to the axes of Figure 8.  These extra fluctuations disappear for the dimer as the length of the lattice becomes large.
Nevertheless, the persistent characteristic corroborating the earlier
work of~\cite{lepri} identified here
concerns the asymmetry of transmission between
incoming from the left ($k_0>0$) and from the right ($k_0<0$).

\section{Conclusions \& Future Challenges}
\label{conc}

In this work, we have  presented a methodology for addressing
the spectral analysis of linear (or linearized) chains that possess
an embedded oligomer defect. The technique, which draws
a series of direct parallels with the approach used in~\cite{malaz},
consists of the solution of the linear problem (with its boundary
conditions) on the lattice in which the defect is embedded.
Subsequently these two linear solutions (on the left and
right of the defect) are used as boundary conditions of
the embedded region. Nontrivial solutions require that
the determinant of a suitably defined matrix vanishes,
ultimately leading to a polynomial problem
for the eigenvalue, of the form given by (\ref{eq: det=0,nu=nu}).
{ For the dimer, the polynomial
problem can be formulated explicitly, see Eq.~\ref{eq: lin2V1=-V2polyn}.
In the general case, the eigenvalues and eigenvectors of the full problem
are obtained by numerical solutions and, the approach can be regarded
as  ``semi-analytic".
In terms of computational cost and efficiency, our approach
is not proposed as an advantageous alternative
to algorithms in which eigenstates are computed by a
computer algebra system directly from the relevant matrix equation.
Its main merit is that of being exact (in as far as its analytical
formulation goes) and of providing an analytical intuition towards
the form of the eigenvectors of relevance to the problem.}

We have applied the approach to genuinely linear problems with embedded
complex oligomers (Section \ref{sec: L}), and, in the second part of the paper, to solve
the stability problem as it arises from linearization around
(analytically known or numerically computed)
extended solutions with an embedded nonlinear defect (section \ref{sec: N}).
Finally, we calculated the spectrum and eigenstates in a number of cases of
interest for \cP\cT symmetric and Hamiltonian potentials.
Our semi-analytical approach shows excellent agreement with numeric
computations of the spectrum.

{One important conclusion that arises from the specific instances that
we have considered above is that}
the instabilities found in the nonlinear cases are generically oscillatory,
with unstable eigenvalues having a non zero real and imaginary part
and thus leading to oscillatory growth. { As intutively
expected, the unstable eigenvectors are exponentially localized around the defects,
as only perturbations located on nonlinear sites can grow.}
The question, however, of the dynamical development
of the instability and of the ultimate fate of such solutions
is still in many respects open to more systematic studies.
In the Hamiltonian case, there is numerical evidence that the instability
leads to self-trapping of energy at the oligomer \cite{unp}.
On the other hand, the dynamics of the $\cP \cT$ case is fundamentally
different
in the instability development, usually ending in an indefinite growth
at the gain site~\cite{ourselves}.

{  As formulated, the
approach is fairly general and applies to the spectral
problem of arbitrary stationary solutions of finite segments embedded
in linear chains.
This is true not only in the more standard Hamiltonian
case of structural defects that has been investigated even
experimentally -as regards its defect states and their interaction
with discrete solitons~\cite{roberto}--; it is also true more
generally for the open  $\cP \cT$-symmetric system case considered
herein. In the latter, experiments have recently tracked the
dimer problem~\cite{dnc2}, although longer size oligomer problems
have not been experimentally reported  as of yet.
An especially interesting generalization would be to extend
the present approach to higher dimensional settings}. A
priori, the solution of the linear problem is available in this
case as well, e.g. for a square defect. Yet the matching
of the solutions from the four relevant directions in the two-dimensional
case presents considerable challenges. Generalizing the approach
for both linear and nonlinear problems in such a case would be
a particularly interesting topic for future studies. Another interesting
issue to examine, even in the one dimensional setting, is the effect
of boundary size on the eigenvalues of the problem (especially, so
the $\cP \cT$-symmetric one). Both our earlier work~\cite{ourselves} and
that of others~\cite{sukh} have suggested interesting effects stemming
from the finite size of the domain (and the associated
boundary conditions) within which the embedded defect
lies. Studying such effects and their scaling over the domain size would
be another interesting direction for future work. Relevant themes
are currently under consideration and will be reported in future
publications.



\begin{thebibliography}{99}


\bibitem{tsir1} G.P. Tsironis, M.I. Molina and D. Hennig,
Phys. Rev. E {\bf 50}, 2365 (1994).\\

\bibitem{moli} M.I. Molina, H. Bahlouli
Phys. Lett. A {\bf 284}, 87 (2002).\\

\bibitem{efrem} K. Hizanidis, Y. Kominis and N.K. Efremidis,
Opt. Express {\bf 16}, 18296 (2008).\\

\bibitem{kovalev} P. G. Kevrekidis, Yu. S. Kivshar, and A. S. Kovalev
Phys. Rev. E 67, 046604 (2003)\\

\bibitem{Hennig99}
G.P. Tsironis and D.~Hennig.
\newblock {\em Phys. Rep.}, {\bf 307}, 333 (1999).\\

\bibitem{davydov} A. S. Davydov, \emph{Solitons in Molecular Systems},
2nd ed. (Reidel, Dortrecht, 1991).\\

\bibitem{holstein} T. Holstein, Ann. Phys. (N.Y.) {\bf 8}, 325 (1959);
see also e.g. G.D. Mahan, {\it Many-Particle Physics} (Plenum, New York, 1981).\\

\bibitem{Scott2003} A. C. Scott,
  \emph{Nonlinear science. Emergence and dynamics of coherent
  structures.} (Oxford University Press, Oxford, 2003).\\

\bibitem{kenkre} V. M. Kenkre and D. K. Campbell,
Phys. Rev. B {\bf 34}, 4959–4961 (1986). \\

\bibitem{molina1} M.I. Molina and G.P. Tsironis,
Phys. Rev. B {\bf 47}, 15330 (1993).\\

\bibitem{molina2} D. Chen, M.I. Molina, G.P. Tsironis,
J. Phys.: Condens. Matter {\bf 5}, 8689 (1993).\\

\bibitem{malaz} B.~A. Malomed and M.~Ya. Azbel.
Phys. Rev. B, {\bf 47}, 10402--10406 (1993).\\

\bibitem{BMalDing} B. A. Malomed, E. Ding, K. W. Chow and S. K. Lai, Phys. Rev. E {\bf 86}, 036608 (2012).\\

\bibitem{BMalBrazhnyi} V. A. Brazhnyi and B. A. Malomed, Phys. Rev. A {\bf 86} 013829 (2012).\\

\bibitem{BMalBrazhnyiSSM} V. A. Brazhnyi and B. A. Malomed, Phys. Rev. A {\bf 83} 053844 (2011).\\

\bibitem{SolPinHS} C. H. Tsang, B. A. Malomed, C. K. Lam and K. W. Chow, Eur. Phys. J. D {\bf 59}, No. 1, 81-89 (2010).\\


\bibitem{Miroshnichenko2009}
A.~E. Miroshnichenko.
\newblock {\em Phys. Lett. A}, 373, 3586, (2009).\\


\bibitem{Tietsche2008}
S.~Tietsche and A.~Pikovsky.
\newblock {\em Europhys. Lett.}, 84:10006 (2008).\\


\bibitem{pgk} K. Li and P. G. Kevrekidis
Phys. Rev. E {\bf 83}, 066608 (2011)\\

\bibitem{Delyon86}
F.~Delyon, Y.E. L{\'e}vy, and B.~Souillard,
Phys. Rev. Lett., \textbf{57}, 2010--2013 (1986).\\

\bibitem{Wan90}
Y.~Wan and CM~Soukoulis, Phys. Rev. A, \textbf{41}, 800--809 (1990).\\

\bibitem{Li96}
Q.~Li, CT~Chan, K.M~Ho, and C.M.~Soukoulis,
Phys. Rev. B, \textbf{53} 15577--15585, (1996).\\

\bibitem{R1} C. M. Bender and S. Boettcher,
Phys. Rev. Lett. {\bf 80}, 5243-5246
(1998).\\


\bibitem{R2} C. M. Bender, S. Boettcher and P. N. Meisinger,
J. Math. Phys. {\bf 40},
2201-2209 (1999).\\




\bibitem{R21} C. M. Bender,
Rep. Prog. Phys. {\bf 70}, 947-1018 (2007).\\


\bibitem{dnc1} A. Guo,
G. J. Salamo, D. Duchesne, R. Morandotti, M. Volatier-Ravat, V. Aimez, G. A. Siviloglou, and D. N. Christodoulides,
Phys. Rev. Lett. {\bf 103}, 093902 (2009).\\

\bibitem{dnc2} C.E. R{\"u}ter,
K.G. Makris, R. El-Ganainy, D.N. Christodoulides, M.Segev, and D. Kip,
Nature Physics {\bf 6}, 192-195 (2010).\\

\bibitem{dnc_nat} A. Regensburger,
C. Bersch,
M.-A. Miri,
G. Onishchukov,
D.N. Christodoulides
and U. Peschel,
Nature     {\bf 488},
    167-171 (2012).\\


\bibitem{lepri} S. Lepri and G. Casati, Phys. Rev. Lett.
{\bf 106}, 164101 (2011).\\


\bibitem{kosevich} Yu. A. Kosevich, Phys. Rev. B,
{\bf 52}, 1017 (1995).\\

\bibitem{chiara1} N. Boechler, G. Theocharis and C. Daraio,
Nature Materials {\bf 10}, 665 (2011).\\


\bibitem{l5} M. Scalora, J. P. Dowling, C. M. Bowden, and M. J.
Bloemer, J. Appl. Phys. {\bf 76}, 2023 (1994);
M. D. Tocci, M. J. Bloemer, M. Scalora, J. P. Dowling, and
C. M. Bowden, Appl. Phys. Lett. {\bf 66}, 2324 (1995).\\


\bibitem{Kevrekidis}
P.~G. Kevrekidis.
\newblock {\em The Discrete Nonlinear Schr\"odinger Equation}.
\newblock Springer Verlag, Berlin, 2009.\\


\bibitem{ourselves} J.~D'Ambroise, P.G. Kevrekidis, and S.~Lepri.
J. Phys. A {\bf 45}, 444012 (2012).\\

\bibitem{unp} S. Lepri and G. Casati, preprint arXiv:1211.4996.\\

\bibitem{roberto} R. Morandotti, H. S. Eisenberg, D. Mandelik, Y. Silberberg, D. Modotto, M. Sorel, C. R. Stanley, and J. S. Aitchison, Opt. Lett. {\bf 28},
834 (2003).\\

\bibitem{sukh} A.A. Sukhorukov, S.V. Dmitriev, S.V. Suchkov and Yu.S.
Kivshar, Opt. Lett. {\bf 37}, 2148 (2012).\\


\bibitem{footnote1} This imposition is
for simplicity/specificity of the calculation; other boundary conditions
can be treated in similar ways by applying them to the solution
of Eq.~(\ref{eq: Lansatz}), to obtain two conditions relating $A$, $B$, $C$ and
$D$.\\

\bibitem{footnote2} It should be
noted that $\gamma_c=1$ for the pure dimer (not embedded on a lattice)
and for the dimer embedded on a finite lattice. Remarkably, the
case of a dimer embedded in the infinite lattice has a different
$\gamma_c=\sqrt{2}$ (D.E. Pelinovsky, personal communication).
However, as our aim herein is to analyze the finite lattice case, we
do not focus on this aspect further in what follows.


\end{thebibliography}
\end{document}